\documentclass[%
 reprint,
preprintnumbers,
 amsmath,amssymb,
 aps,
]{revtex4-1}

\usepackage[pdftex]{graphicx,color}
\usepackage{graphicx}% Include figure files
\usepackage{dcolumn}% Align table columns on decimal point
\usepackage{bm}% bold math
\usepackage{subfigure}
\begin{document}

\preprint{OU-HET-1000}

\title{Chaos of QCD string from holography} 
% Force line breaks with \\

\author{Tetsuya Akutagawa}
\author{Koji Hashimoto}%
\author{Keiju Murata}%
\author{Toshihiro Ota}%
\affiliation{%
 Depertment of Physics, Osaka University, \\
  Toyonaka, Osaka 560-0043, Japan
}%


%\affiliation{
%\affiliation{
%\affiliation{%



\begin{abstract}
It is challenging to quantify chaos of QCD, because non-perturbative QCD accompanies
non-local observables.
By using holography, we find that QCD strings at large $N_c$ and strong coupling limit 
exhibit chaos, and measure their Lyapunov exponent at zero temperature.
A pair of a quark and an antiquark separated by $L_q$ in the large $N_c$ QCD is dual
to a Nambu-Goto string hanging from the spatial boundary of the D4-soliton geometry.
We numerically solve the motion of the string after putting a pulse force on its boundaries.
The chaos is observed for the amplitude of the force larger than a certain lower bound.
The bound increases as $L_q$ grows, and its dependence is well approximated by a hypothesis
that the chaos originates in the endpoints of the QCD string.

%To explore non-perturbative dynamics of QCD, we study full nonlinear motion of an open string in the D4-soliton background. 
%In general, the motion of a string in a curved background shows chaotic behavior. 
%Through the AdS/CFT, the chaos of an open string hanging on the D4 boundary is translated into chaos of Wilson loops, i.e. force acting on quarks in the dual gauge theory. 
%We found that in large $N_{c}$ pure Yang-Mills theory force acting on quarks is generically sensitive to initial perturbations. 
%In addition, we obtain a phase diagram of the chaos, which implies that for larger quark distance the chaos diminishes. 
%We also give a direct numerical simulation that a static Wilson loop through D4-soliton geometry realizes a $q\bar{q}$ potential which has both Coulombic phase and confining phase. 
%Our results suggest that time evolution of observables in QCD may be generically chaotic and 
%especially provide a theoretical prediction for time evolution of the Wilson loop. 
%\begin{description}
%\end{description}
\end{abstract}

\pacs{Valid PACS appear here}% PACS, the Physics and Astronomy
\maketitle

%\tableofcontents

\section{\label{sec:intro}Introduction: QCD chaos}

How chaotic is QCD? --- a question which is simple but unanswered, should drive
the understanding of our universe based on quantum field theories.
It is challenging to define the extent of chaos for QCD, because QCD is 
truly quantum while the popular measure of chaos, the Lyapunov exponent, 
is defined classically. Analyses based on weakly coupled picture 
\cite{Matinyan:1981dj,Matinyan:1986nw,Muller:1992iw,Biro:1994sh,Gong:1993xu,Kunihiro:2010tg,Muller:2011ra,Iida:2013qwa,Tsukiji:2016krj}
and on out-of-time ordered correlators \cite{Maldacena:2015waa} (which define a quantum chaos) 
suggest a QCD chaos at high temperature,
but what about the usual picture of hadronic phase of QCD?

As lattice QCD, a popular strategy to study non-perturbative nature of QCD,
still lacks a way to follow time dependence necessary to
analyze any chaos, we need some other way.
The holography, or the AdS/CFT correspondence \cite{Maldacena:1997re}, 
is suitable for the purpose.
Taking a large $N_c$ limit and 
a strong coupling limit, QCD is approximated by a classical gravity dual,
while keeping the quantum nature and the time dependence of QCD. 
In this paper, we analyze chaos of a quark antiquark pair by using the holography. 
We find a condition for the chaos to occur, and draw a phase diagram of the QCD chaos.

The study of chaos in the AdS/CFT was initiated in \cite{Zayas:2010fs}.
While the chaos of chiral condensate in QCD was studied in \cite{Hashimoto:2016wme,Akutagawa:2018yoe} 
via the holography, 
the physical excitation of QCD at low energy is non-local. Wilson loops, and
pairs of quark-antiquark connected by an open Wilson loop, are the low energy physical degrees
of freedom of QCD and their quantization provides the hadronic world. Spectra of hadrons 
exhibit quantum chaos \cite{Pascalutsa:2002kv}, so, we need to locate the origin of the chaos of QCD,
and measure the extent of the QCD chaos, based on the non-local Wilson loops.

Here we have to remind the readers of the fact that a Nambu-Goto (NG) 
closed string in three spatial dimensions,
a phenomenological model of glueballs in the large-$N_c$ QCD, is 
integrable. Then, what is the origin of
the QCD chaos? Naively, we can expect two possible origins: one is the boundary of the NG 
string, which is the quark, and the other is the thickness of the QCD string which has not been
taken care of for the three-dimensional NG string. The question can be addressed
in holography, because the QCD string corresponds to a NG string in the higher-dimensional
spacetime in the gravity dual,
and its static nature, such as the quark boundaries and the thickness, has been well-studied. 
We here provide a detailed analysis of a quark antiquark pair in motion, and locate
the origin of the QCD chaos.

Through the AdS/CFT, The Wilson loop in QCD is identified with a NG string 
\cite{Maldacena:1998im, Rey:1998ik} hanging down 
from the boundary of confining geometry \cite{Witten:1998zw} which is considered to 
be a dual to a pure 4-dimensional Yang-Mills theory. The $q\bar{q}$
potential is the free energy of the string. Since the geometry has the bottom of the spacetime,
the hanging string has the part sitting at the bottom, which provides the QCD string tension,
and the parts connecting the bottom and the boundary, which correspond to the quarks in a 
gluon cloud. The motion of the QCD string is caused by a pulse force acting on the infinitely massive 
quarks, and we solve numerically the motion of the NG string in the geometry.
The chaotic Lyapunov exponent is observed when the strength of the pulse force exceeds a certain bound. We study the dependence of the bound on the interquark distance, 
and find that the $q\bar{q}$ pair is less chaotic for larger interquark distances.

Our numerical result
is explained well by a popular effective picture of the quarks connected by a long QCD string, assuming that 
the QCD string motion is integrable while the endpoint regions (the quarks with a gluon cloud) are chaotic.
It suggests that the chaos of QCD string originates in its endpoints.
The chaos of motion of closed NG string in various geometry has been studied \cite{Basu:2011di,Basu:2012ae,PandoZayas:2012ig} (see also \cite{Basu:2013uva,Giataganas:2013dha,Giataganas:2014hma,Bai:2014wpa,Asano:2015qwa,Ishii:2015wua,Ishii:2015qmj,Basu:2016zkr}),
which corresponds to the chaos due to the thickness of the QCD string. Our study about
the quark antiquark pair shows a different origin of the QCD chaos.

The organization of this paper is as follows. First, in Sec.~\ref{sec:static},
we review the static Wilson loop in the holographic QCD, and introduce our coordinate system in
the bulk. In Sec.~\ref{sec:rect}, a rectangular NG string is introduced as a toy model,
and its fluctuation analysis is presented to show the existence of the chaos and the chaos energy 
bound. Our main numerical study of the NG string in motion in holography is presented in Sec.~\ref{sec:chaos}. There we find Lyapunov exponent of the string motion, 
and draw a phase diagram of chaos, as a function of $\epsilon$ which is the magnitude of the
pulse force and $L_q$, the interquark distance. The chaotic behavior is observed in
the interquark force which is an observable of QCD.
Sec.~\ref{sec:model} is for a discussion to locate the origin of chaos. We introduce a
simple effective picture of an open QCD string and discuss its chaos, to fit the numerical result of the
phase diagram obtained in Sec.~\ref{sec:chaos}. We conclude that the chaos originates in the
endpoints of the QCD string, the quarks.
Sec.~\ref{sec:summary} is for a summary and discussions.
App.~\ref{sec:num} provides details of numerical calculations. 
App.~\ref{sec:force} calculates the formula for the interquark force.

\section{\label{sec:static}Confining geometry and $q\bar{q}$ potential}
%static string

In this section, we first review the confining geometry \cite{Witten:1998zw} 
and compute the static $q\bar{q}$ potential through the AdS/CFT, based on the dictionary 
\cite{Maldacena:1998im, Rey:1998ik}. The string configuration serves as an initial one upon which
an external pulse force is put to produce a time-dependent motion of the QCD string, later in
Sec.~\ref{sec:chaos}.

The D4-soliton background holographically corresponds to a five dimensional super Yang-Mills theory
on a non-supersymmetric circle, giving a four-dimensional pure Yang-Mills theory
at low energy \cite{Witten:1998zw}. 
The background is an example of confining geometries which has the bottom of the spacetime. 
Let us first obtain the static NG string configuration hanging down from the boundary of the spacetime, to calculate the expectation value of the Wilson loop in the Yang-Mills theory.
%We show this geometry has a confining phase by direct numerical calculation of the Wilson loop.

The D4-soliton background is of the following form \cite{Witten:1998zw,Kruczenski:2003uq}:
\begin{align}
\label{metric1}
ds^2&=\left( \frac{U}{\mathcal{R}}\right)^{3/2} (\eta_{\mu \nu}dx^{\mu}dx^{\nu}+f(U)d\tau^2) \nonumber \\
&\qquad+\left( \frac{\mathcal{R}}{U} \right) ^{3/2} \frac{dU^2}{f(U)}+\mathcal{R}^{3/2}U^{1/2} d\Omega_4^2, \\
%e^{\phi}&=g_s \left( \frac{U}{\mathcal{R}}\right) ^{3/4}, \, F_4 = \frac{2\pi N_c}{V_4}\epsilon_4, \, 
&f(U)=1-\frac{U_{\text{KK}} ^3}{U^3}. 
\end{align}
The coordinates $x^{\mu}$ and $\tau$ are the directions along the D4-branes, and the $\tau$ direction is compactified on $S^{1}$. 
%$d\Omega_4$ and $\epsilon_4$ are the line elements and the volume form respectively, which are invariant under $SO(5)$ transformation. 
The coordinate $U$ is a radial direction transverse to the D4-branes.
% and $U_{\text{KK}}$ is a parameter. $\mathcal{R}$ is related to string constants by $\mathcal{R}^3=\pi g_s N_c l_s ^3$. 
To avoid a conical singularity at $U=U_{\text{KK}}$, the period of the $\tau$ direction must be
\begin{align}
\delta \tau = \frac{4\pi}{3} \frac{\mathcal{R}^{3/2}}{U_{\text{KK}} ^{1/2}}=\frac{2\pi}{M_{\text{KK}}}, 
\end{align}
so $1/M_{\text{KK}}$ is the radius of $S^{1}$. 
Parameters in the metric can be expressed by those of the dual gauge theory as 
%The parameters $\mathcal{R},U_{KK}$ and $g_s$ are related to the gauge theory parameters by 
\begin{align}
\mathcal{R}^3 = \frac{1}{2}\frac{g_{\text{YM}}^2 N_c l_s ^2}{M_{\text{KK}}}, \; U_{\text{KK}}= \frac{2}{9} g_{\text{YM}}^2 M_{\text{KK}} l_{s}^2.
%, \, g_s =\frac{1}{2\pi} \frac{g_{\text{YM}}^2}{M_{\text{KK}} l_s}.
\end{align}
The motion of the NG string studied in Sec.~\ref{sec:chaos} often goes through
the tip of the geometry $U=U_{\rm KK}$, thus we need a coordinate system which does not
have a coordinate singularity there.
The new coordinate $r$ is introduced as
%For later use, we would like to change the coordinate into a form suitable for numerical calculation. 
%We introduce a coordinate $r$ related to $U$ with
\begin{eqnarray}
U(r) = U_{\text{KK}} (1+ \tan^2 r) .
\end{eqnarray}
Then, the metric becomes
\begin{align}
ds^{2} = \frac{4}{3}\lambda l_{s}^{2} \frac{1}{\cos^{3}r} \left[ \frac{M_{\text{KK}}^{2}}{9}(-dt^{2} \!+\! d\vec{x}^{2}) + \frac{dr^{2}}{1\!+\!\cos^{2}r \!+\! \cos^{4}r}  \right], \label{eq:d4metricr}
%ds^2&=&\left(\frac{U_{kk} \sec ^2(r)}{\mathcal{R}}\right)^{3/2}\left(\eta_{\mu \nu}dx^{\mu}dx^{\nu}+(1- \cos^6(r))d\tau^2\right)\\
%&&+\frac{4  \mathcal{R} U_{kk} \sec^8(r) \sqrt{\frac{\mathcal{R} \cos^2(r)}{U_{kk}}}}{\tan ^4(r)+3 \tan^2(r)+3}dr^2+\mathcal{R}^{3/2} \sqrt{U_{kk} \sec ^2(r)}d\Omega_4^2 .
\end{align}
where $\lambda=g_{\text{YM}}^{2}N_{c}$ is the 't Hooft coupling. 
Since we are interested in QCD we do not consider $\tau$ and $\Omega_{4}$ directions in the following, and so here we have omitted them. 
In this coordinate, the bottom of the D4-soliton $r=0$ is totally regular. 
%while in original coordinate $U=U_{\text{KK}}$ was a coordinate singularity. 
The asymptotic boundary of D4-soliton is now $r=\pi/2$, which was $U\to \infty$ in the $U$ coordinate.

The $q\bar{q}$ potential is given by the energy of a static NG open string in the geometry of
the gravity dual
\cite{Maldacena:1998im, Rey:1998ik}. 
We consider a Wilson loop with the quark-antiquark separation $L_q$, and take an ansatz that the string is extended in the $x^1$-$r$ plane and the endpoints of the string are located at $x^1 = \pm L_q/2$. 
The NG action in the geometry \eqref{eq:d4metricr} is
\begin{eqnarray}
S_{\text{NG}}=-\frac{1}{2 \pi \alpha'}\int d\tau d\sigma \sqrt{-h},
\end{eqnarray}
where $h=\det (h_{ab})$ and $h_{ab}$ is an induced metric on the worldsheet. 
We take the static gauge: $(\tau,\sigma)=(t,r)$ and then the static solution is provided as 
$x^1 =X_1(r)$.\footnote{
We will use small letters for target space coordinates  
and 
capital letters for functions specifying the string position.
}
For numerical calculations, we choose a unit 
%{\color{red}{$U_{\text{KK}}=1,R=1$ and then} }
$M_{\text{KK}} = 3/2$. 
The NG action now becomes
\begin{align}
\label{action}
S_{\text{NG}} &= \frac{\lambda}{6\pi} \mathcal{T}
\!\!  \int_{\pi/2}^{r_{\rm center}} \!\!\!\!\!
dr \frac{1}{\cos^{3}r} \left[ \big( \vec{X}' \big)^{2} \!\! 
+ \frac{4}{1+\cos^{2}r + \cos^{4}r} \right]^{1/2} 
%S=-\frac{1}{2\pi\alpha'}\int dtdr \frac{\sqrt{2}}{18}  \sqrt{\frac{ \sec ^{10}(r) \left((9/4) (8 \cos(2 r)+\cos(4 r)+15) X_1'(r)^2+72 \right)}{\tan^4(r)+3 \tan ^2(r)+3}}
\end{align}
where $' = \partial_{r}$ and $\mathcal{T}=\int dt$. 
The integration region is $\pi/2 \leq r \leq r_{\rm center}$ where $r=r_{\rm center}$ is
the point of the bottom of the hanging string, which should solve the equation $X(r_{\rm center})=0$ due to
the parity symmetry $X(r) =-X(-r)$ following from our boundary condition.
Solving the equations of motion, we obtain static configurations of string for each quark separations $L_{q}$. 
%So the equation of motion becomes
%\begin{align}
%\label{eom}
%&&-\frac{1}{12  A(r)  \sqrt{\sec ^6(r) \left(X'(r)^2+\frac{32 }{A(r)}\right)} \left(A(r) X'(r)^2+32\right)}  \biggl[\sec ^7(r) (64 (38 \cos(r)+9 \cos(3 r)+\cos(5 r)) X''(r) \nonumber \\
%&&+64 (58 \sin(r)+11 \sin (3 r)+\sin (5 r)) X'(r)+12 \sin (r)  (8 \cos(2 r)+\cos(4 r)+15)^2  X'(r)^3)\biggr]=0
%\end{align}
%where
%\begin{eqnarray*}
%A(r) \equiv 8 \cos(2 r)+\cos(4 r)+15.
%\end{eqnarray*}
%Figure
\begin{figure}[tb]
 \begin{center}
  \includegraphics[width=55mm, trim=20 10 10 0]{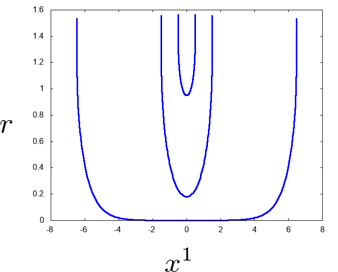}
 \end{center}
 \caption{Static strings in the D4-soliton background. From the innermost string, $L_{q}=1,3,12$.}
 \label{wilsonloop}
\end{figure}
Fig.~\ref{wilsonloop} shows the configurations of the static string in the $x^1$-$r$ plane. 
When the quark-antiquark separation becomes larger, the tip of hanging string sticks to 
the bottom of the geometry $r=0$.
This actually implies that confining potential appears.

Let us evaluate the $q\bar{q}$ potential holographically.
%We calculate on-shell action. 
Considering the on-shell NG action $S_{\text{NG}}[\tilde{X}]$, where $\tilde{X}(r)$ is a static solution to the equation of motion, 
%action(\ref{action})
%\begin{eqnarray}
%S=-\frac{T}{2\pi\alpha'}\int dr \frac{\sqrt{2}}{18}  \sqrt{\frac{ \sec ^{10}(r) \left((9/4) (8 \cos(2 r)+\cos(4 r)+15) \hat{X}_1'(r)^2+72 \right)}{\tan^4(r)+3 \tan ^2(r)+3}}
%\end{eqnarray}
%where T is the constant of time integration. 
the $q\bar{q}$ potential is given by %{\color{red}{(The factor 2 correct?)}}
\begin{align}
E=-\frac{2}{\mathcal{T}} S_{\text{NG}}[\tilde{X}] .
\end{align}
This has a divergence which stems from the infinitely long string hanging from the boundary, but it can be naturally understood as the infinite quark mass. 
Subtracting the contribution of that, the $q\bar{q}$ potential turns out to be $E-E_0 
%-\frac{2}{2\pi\alpha'}\int dr \biggl[\frac{\sqrt{2}}{18}  \sqrt{\frac{ \sec ^{10}(r) \left((9/4) (8 \cos(2 r)+\cos(4 r)+15) \hat{X}_1'(r)^2+72 \right)}{\tan^4(r)+3 \tan ^2(r)+3}}-1\biggr]
$, where  %{\color{red}{(Meaning of this expression?)}}
\begin{align}
E_0 = -\frac{2}{\mathcal{T}}
\left(
\frac{\lambda}{6\pi} \mathcal{T}\int_{\pi/2}^0 
dr \frac{1}{\cos^{3}r} \frac{2}{\sqrt{1+\cos^{2}r + \cos^{4}r} }\right).
%E_0 = -\frac{2}{2\pi \alpha'}\int_{\pi/2}^{0} dr.
\end{align}
The quantity in the parenthesis is \eqref{action} with $X'=0$, except that the integration region is
$\pi/2 \leq r \leq 0$.
%Figure
\begin{figure}[tb]
 \begin{center}
  \includegraphics[width=55mm, trim=20 10 20 10]{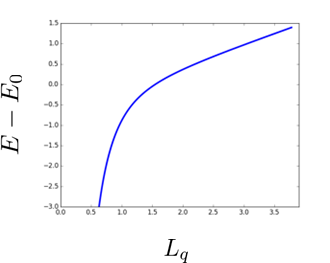}
 \end{center}
 \caption{The relation between quark-antiquark separation $L_{q}$ and $q\bar{q}$ potential.}
 \label{potential}
\end{figure}
Fig.~\ref{potential} shows the relation between the quark-antiquark separation $L_q$ 
and the $q\bar{q}$ potential $E-E_0$. 
When $L_{q}$ is large, the potential becomes linear in $L_q$, 
%If the quark-antiquark separation is small, the potential become the five dimensional Coulomb's potential. 
which means it is a confining potential. 
This is totally consistent with the study developed in \cite{Brandhuber:1998er,Greensite:1998bp,Kinar:1998vq}.

%So far, we have confirmed that the string in the D4 background realizes confining potential via the AdS/CFT just by using static string. 

In the next section, we consider a toy model of the motion of the string and study a chaos bound,
and in Sec.~\ref{sec:chaos}, 
we investigate numerically the full time-dependent 
dynamics of the string in the D4-soliton background and interquark force in the gauge theory.

%%%%%%%%%%%%%%%%%%%%%%%%%%%%%%%%%%%%%%%%%%%%%%%%%
\section{\label{sec:rect}Toy model of string in motion}

\begin{figure}[tb]
 \begin{center}
  \includegraphics[width=85mm]{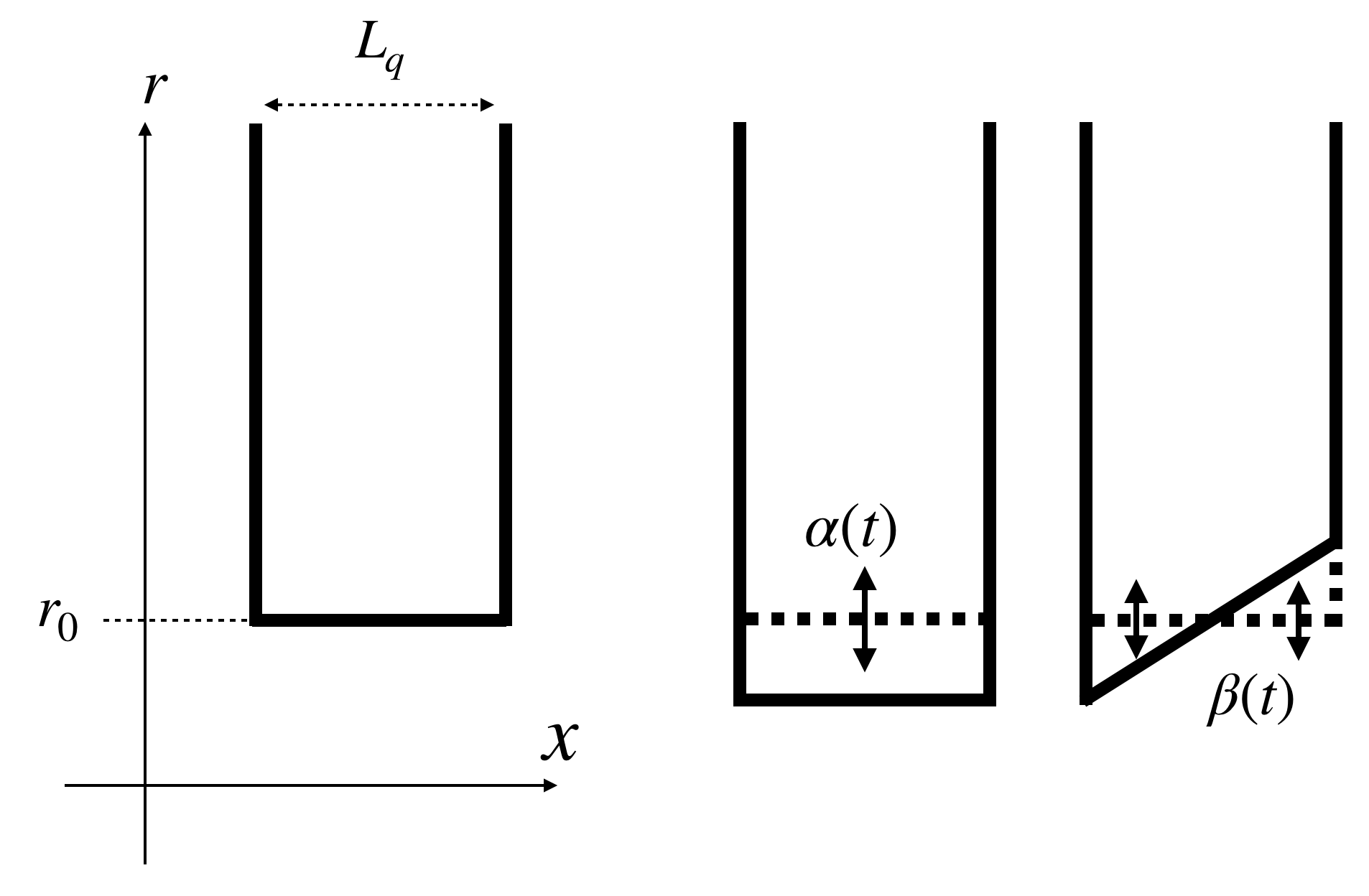}
 \caption{Left: the toy-model string, of the rectangular shape. Right: the lowest 
 two fluctuation modes we consider, $\alpha(t)$ and $\beta(t)$.}
 \end{center}
 \label{fig:rect}
\end{figure}

Before getting into the full numerical simulation of the NG string in motion, we here first study
the motion of a toy model string to look intuitively how the chaos shows up in the motion.
The toy model assumes the shape of the string and its fluctuation modes: the shape of the
toy string is rectangular, and the fluctuation modes are only of two types,
one is the motion keeping the rectangular shape, and the other is the motion giving a linear slope
at the bottom of the rectangular string, see Fig.~\ref{fig:rect}.
As is seen in comparison to the actual shape in Fig.~\ref{wilsonloop}, this toy model could capture
some intrinsic feature of the motion of the NG string. In fact, the rectangular string hanging down from
the boundary has been used in many literature to mimic the QCD string holographically, and 
it was used in \cite{Hashimoto:2018fkb} to show a universal chaos behavior near black hole horizons.

We here show that the toy model has no chaos when the total energy of the string is small, while
for a larger energy the chaos appears. We estimate the lower energy bound of the chaos in the model,
and the bound is used in the next section
for intuitively understanding the simulation results.
The existence of the chaos bound itself is easy to understand: For a very small fluctuation around the
static string shape, the motion is that of a harmonic oscillator, so there is no chaos. On the other hand,
if one puts a larger energy, any modes are excited and interacting with each other, generally inducing chaos.
We are interested in the energy lower bound and its dependence on the distance between the quarks, $L_q$.
 
To obtain the fluctuation action of the toy string model, first we determine the static stable configuration.
Denoting the location of the bottom of the rectangular 
string as $r=r_0$ with $-L_q/2\leq x \leq L_q/2$, the NG action is
\begin{align}
S_{\rm NG} = & \frac{-2\lambda}{27\pi} {\mathcal T} M_{\text{KK}}
\left[
\frac{L_q M_{\text{KK}}}{\cos^3 r_0} 
\right.
\nonumber \\
&\left.
+ 6 \int_{r_0}^{\pi/2} \!\!\!\frac{dr}{\cos^3r\sqrt{1+\cos^2r + \cos^4r}} 
\right] \, .
\end{align} 
Extremizing of this action with respect to $r_0$, we obtain the relation between $r_0$ and $L$ for the static
stable rectangular string,
\begin{align}
L_q M_{\text{KK}} = \frac{2 \cos r_0}{\sin r_0 \sqrt{1+\cos^2r_0 + \cos^4r_0}} \, .
\end{align}
In particular, for a large interquark distance $L_q M_{\text{KK}}\gg 1$, this relation is rephrased as
\begin{align}
r_0 = \frac{2}{\sqrt{3}}\frac{1}{L_q M_{\text{KK}}}+ \frac{4}{9\sqrt{3}}\frac{1}{(L_q M_{\text{KK}})^3} + \cdots \, .
\end{align}
Let us proceed to obtain the fluctuation action. We include the lowest two modes, which are represented by the
following linear shape of the bottom of the string,
\begin{align}
r = r_0 + \alpha(t) + \beta(t)\frac{x}{L_q} \quad (-L_q/2\leq x \leq L_q/2)
\end{align}
The fluctuation modes $\alpha(t)$ and $\beta(t)$ deform the bottom of the string. 
With this shape it is straightforward to obtain the NG action,
\begin{align}
S = \frac{-2\lambda M_{\text{KK}}}{9\pi} \int dt \left( {\cal L}_{\text{radial}} + {\cal L}_{\text{bottom}}\right)
\label{totalS}
\end{align}
where the radial and the bottom parts of the action of 
the string are given by
\begin{align}
{\cal L}_{\text{radial}} = 
& \int_{r_0+\alpha-\beta/2}^{\pi/2} 
\frac{d\sigma}{\cos^3\sigma \sqrt{1+\cos^2\sigma + \cos^4\sigma}} 
\nonumber \\
&+\int_{r_0+\alpha+\beta/2}^{\pi/2}
\frac{d\sigma }{\cos^3\sigma\sqrt{1+\cos^2\sigma + \cos^4\sigma}} \, ,
\end{align}
and
\begin{align}
& {\cal L}_{\text{bottom}} = \frac{M_{\text{KK}}}{3}\int_{-L/2}^{L/2} d\sigma \, \frac{1}{\cos^3 r}
\nonumber \\
&
\times \left[
1-\frac{9}{M_{\rm KK}^2}\left(
\dot{\alpha} + \dot{\beta}\frac{\sigma}{L}
\right)^2 \frac{1}{1+\cos^2r + \cos^4r}
\right]^{1/2}
\nonumber \\
&
\times \left[
1+ \frac{9}{L_q^2 M_{\text{KK}}^2} \beta^2 \frac{1}{1+\cos^2r + \cos^4r}
\right]^{1/2}
\end{align}
with $r = r_0 + \alpha + \beta \sigma/L$ substituted for the last expression.
We expand the total action \eqref{totalS}
to the third order in the fluctuations $\alpha(t)$ and $\beta(t)$. 
The result is
\begin{align}
S = & \; \int \! dt \left[
\text{const.} +a_{11} \dot\alpha^2 + a_{22} \dot\beta^2 - V\right]
\, , 
\\
V \equiv & \;   b_{11} \alpha^2 + b_{22} \beta^2 
\nonumber
\\
&
+ a_{111} \alpha^3 + a_{122} \alpha \beta^2 
\nonumber
\\
&
+ 
b_{111} \alpha \dot{\alpha}^2 + b_{212} \beta \dot{\alpha}\dot\beta + b_{122} \alpha \dot\beta^2 \, .
\end{align}
The coefficients $a$'s and $b$'s are functions of $r_0$, namely, of $L_q$. 
The ``potential'' term $V(t)$ in general includes time-derivative terms.\footnote{
Even if we make a field redefinition of the form $\alpha \to \alpha + \alpha^2 + \alpha \beta$
and $\beta \to \beta + \alpha \beta + \beta^2$ (which respects the 
$x$-party transformation $(\alpha,\beta) \to (\alpha, -\beta)$, we cannot 
absorb the third order terms including time-derivatives (which are $\alpha \dot{\alpha}^2$, $\beta\dot\alpha\dot\beta$ and $\alpha \dot\beta^2$).}

\begin{figure}[tb]
 \begin{center}
  \includegraphics[width=75mm, trim=20 10 20 10]{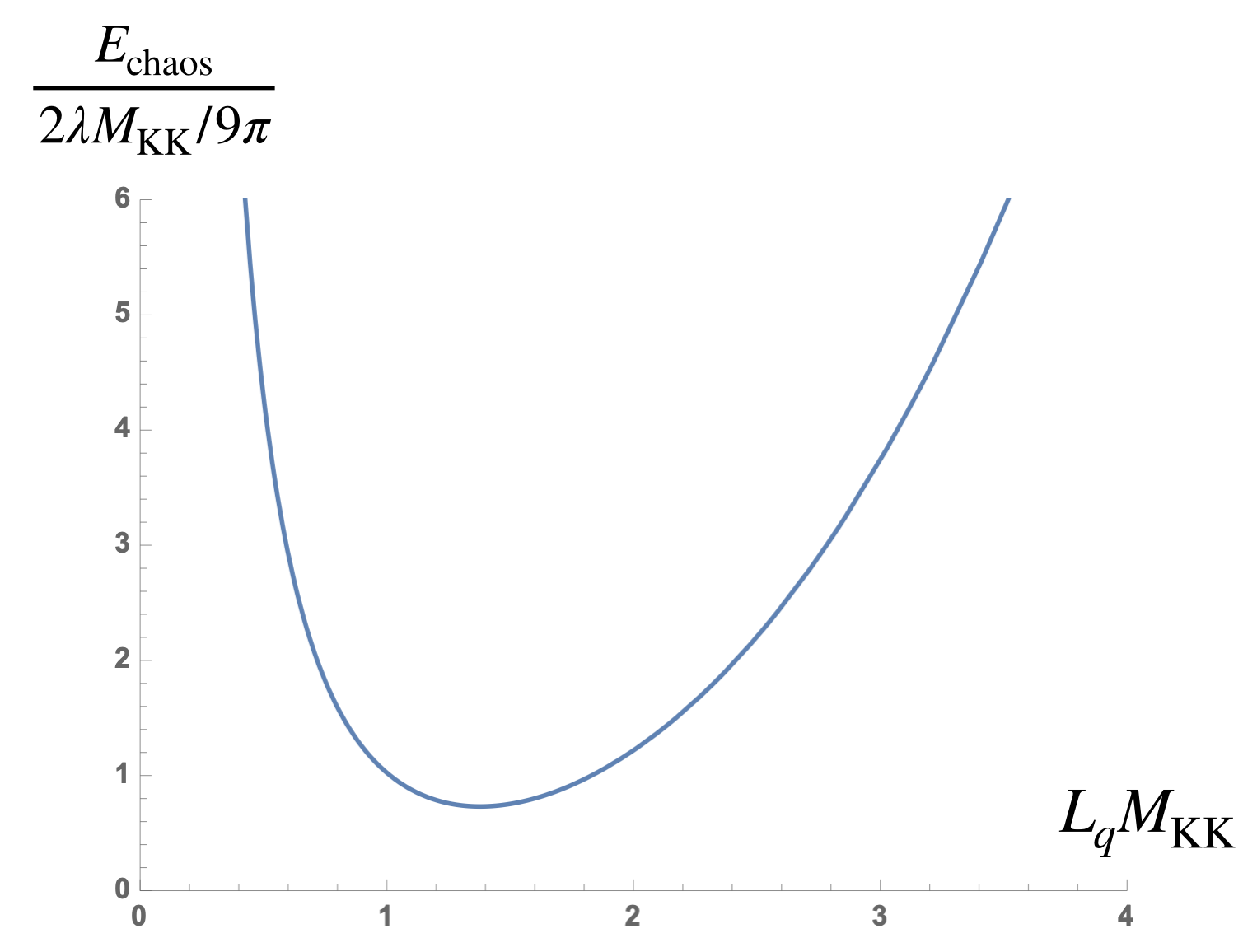}
 \end{center}
 \caption{The energy 
 lower bound \eqref{Echaos} for the chaos in the toy model, as a function of 
 the quark-antiquark separation $L_{q}$.}
 \label{fig:Echaos}
\end{figure}

Generically, for the chaos to occur, the interaction terms (the cubic terms) in $V(t)$
need to contribute. For small fluctuation, only the quadratic terms in $V(t)$ (which are mass
terms for $\alpha(t)$ and $\beta(t)$) provide the full dynamics and it is just a set of 
harmonic oscillators. When the fluctuation is larger, the cubic interaction term contributes,
and the chaos emerges.
To estimate the typical value for the energy lower bound of the chaos to emerge, we pick up
the terms of 
$\alpha(t)^2$ and $\alpha(t)^3$ in $V(t)$ and obtain the energy at which the values of these two terms are equal to each other, under the condition $\dot{\alpha}(t)=0$. This energy is
\begin{align}
E_{\text{chaos}} \equiv 
\frac{2 b_{11}^3}{a_{111}^2}
\label{Echaos}
\end{align}
We plot this chaos bound in Fig.~\ref{fig:Echaos}. We find that the chaos energy bound
diverges for $L_q \to \infty$ or $L_q \to 0$, while it takes its lowest value around 
$L_q \sim 1/M_{\text{KK}}$.
%Figure
This behavior of $E_{\rm chaos}$ is naturally understood, because in the limit
$L_q \to \infty$ or $L_q \to 0$, the system is expected to reduce to an integrable model.
For example, in the former limit $L_q \to\infty$, the string is straight and resides at
the bottom of the D4-soliton geometry, and the coefficients of the 
interaction $V(t)$ is suppressed as $1/L_q$, therefore the chaos disappears. In fact, in this
limit $L_q \gg 1/M_{\rm KK}$ the chaos energy lower bound is calculated as
\begin{align}
E_{\rm chaos} = \frac{3\lambda M_{\rm KK}}{98\pi} (L_q M_{\rm KK})^3 \left(
1 + {\cal O} \left(\frac{1}{(L_q M_{\rm KK})^2}\right)
\right) \, ,
\end{align}
and it diverges as $\sim L_q^3$.

The lessons from the toy model are that the chaos should appear 
at an energy above some nonzero value,
and that the energy lower bound for the chaos diverges as $L_q \to\infty$.
We will find that these are exactly seen in the full numerical simulations presented in
the next section.

%%%%%%%%%%%%%%%%%%%%%%%%%%%%%%%%%%%%%%%%%%%

\section{\label{sec:chaos}Chaos of interquark force}

In this section, we explore the full dynamical motion of string by numerical simulation to examine the chaotic motion. For this purpose, we employ the numerical techniques to study dynamical string developed in \cite{Ishii:2015wua,Ishii:2015qmj}. 
The detail of numerical calculations are summarized in appendix \ref{sec:num}.
To induce the nonlinear dynamics of the string, we instantly move 
the position of the string endpoints at the boundary of the D4-soliton geometry. 
In the gauge theory, this corresponds to an instantly forced motion 
of the quarks: a small deformation of the Wilson loop along the time direction.
%
%In the sense of holography, it turns out to be the 
This produces a 
nonlinear dynamics of the gluon flux tube induced by the motion of quark and antiquark pair, afterwards.

To perform the numerical calculation, we again employ the $r$ coordinate, so the metric is (\ref{eq:d4metricr}). 
Now, as a world-sheet coordinate system, we take double null coordinates $(u, v)$ and specify the string configuration by $t = T(u, v),\, r = R(u, v),\, \vec{x} = \vec{X}(u, v)$. 
The condition on the induced metric $h_{ab}$ for $u, v$ to be null coordinates is given by $h_{uu} = h_{vv} = 0$. 
Then, we have $-h=h_{uv}^2-h_{uu}h_{vv}=h_{uv}^2$.
Thus in the double null coordinate, the Lagrangian for the string is proportional to $h_{uv}$, and working in the unit 
%{\color{red}{$R = U_{\text{KK}} = 1,$}}  
$M_{\text{KK}}=3/2$,
the NG action becomes
\begin{align}
S_{\text{NG}} = -\frac{\lambda}{6\pi}\int dudv \frac{1}{\cos^{3}R}\bigg[ &-T_{,u}T_{,v} +\vec{X}_{,u}\cdot\vec{X}_{,v} \nonumber \\
 &\quad +\frac{4R_{,u}R_{,v}}{1+\cos^{2}R + \cos^{4}R} \bigg]. 
 \label{eq:nullaction}
\end{align}
From this action, we obtain the evolution equations of the string:
\begin{align}
T_{,uv} &= -\frac{3}{2}\tan R \left( T_{,u}R_{,v} + T_{,v}R_{,u} \right), \label{Tuv}\\
R_{,uv} &= -\frac{3}{2}R_{,u}R_{,v} \tan R \bigg[ 1+ \frac{3\cos^{2}R (1+2\cos^{2}R)}{1+\cos^{2}R + \cos^{4}R} \bigg] \nonumber \\
&+\frac{3}{8}\tan R (1+\cos^{2}R + \cos^{4}R) (-T_{,u}T_{,v} + \vec{X}_{,u}\cdot\vec{X}_{,v}), \label{ruv}\\
\vec{X}_{,uv} &= -\frac{3}{2}\tan R \left( \vec{X}_{,u}R_{,v} + \vec{X}_{,v}R_{,u} \right). \label{Xuv}
\end{align}
The double null conditions give constraints
\begin{align}
 C_u&=-T_{,u}^2 +\vec{X}_{,u}^2 +\frac{4R_{,u}^2}{1+\cos^{2}R + \cos^{4}R}=0\ ,\label{conu}\\
 C_v&=-T_{,v}^2 +\vec{X}_{,v}^2 +\frac{4R_{,v}^2}{1+\cos^{2}R + \cos^{4}R}=0\ .\label{conv}
\end{align}
They are conserved by time evolution: $\partial_v (\cos^3 R C_u)=\partial_u (\cos^3 R C_v)=0$.
We impose them at the initial surface and time-like boundaries of the string worldsheet 
and solve time evlution based on Eqs.(\ref{Tuv}-\ref{Xuv}). 
(The numerical technique for solving evolution equations are summarized in  appendix \ref{sec:timeev}.)

Using the residual coordinate transformations, $u\to F(u),\, v\to G(v)$, we put 
the boundaries of the worldsheet at $u-v=0$ and $u-v=\pi$. 
As an initial condition, we take a static string configuration obtained in Sec.~\ref{sec:static}. 
(See appendix \ref{sec:initial} for the detail of the numerical construction of the initial data.)
In the unit 
%{\color{red}{$R = U_{\text{KK}} = 1$ and}} 
$M_{\text{KK}}=3/2$, static string configurations form a one-parameter family of initial conditions for $r_{\text{center}}$, where $r_{\text{center}}$ denotes the initial $r$-coordinate at the tip of the string. 
This $r_{\text{center}}$ is one-to-one correspondent to the interquark distance $L_q$.
Here we use the static solution with the initial condition $r_{\text{center}} = 0.2$ (corresponding to $L_{q} = 2.884$) to demonstrate the simulation of the dynamics.

To induce the nonlinear dynamics of the string, we impose a time-dependent boundary condition on string endpoints. 
Introducing time and spatial coordinates on the worldsheet as $\tau = u+v$ and $\sigma = u-v$, we consider the following forced motion (``quench'') of the string endpoints along the $X_{1}$ direction:
%\footnote{
%We would like to give a comment on our numerical calculation and its relation to some physical 
%implication. 
%In our numerical simulation, we set a tiny cutoff at the endpoints of the string, i.e. $r|_{\sigma=0, \pi} = \pi/2 -c$. 
%If we set exactly $c=0$, the numerical simulation immediately breaks down and we cannot even see regular time evolutions. 
%As long as we take a small enough cutoff $c$, it may not matter to our results. 
%This cut-off may be understood as an integrable deformation of dual CFT's, see
%\cite{Smirnov:2016lqw,Cavaglia:2016oda,McGough:2016lol,Kraus:2018xrn,Chakraborty:2018aji}.
%On one hand, in the context of the AdS/CFT, it has been recently known that a cutoff at the AdS boundary seems to have a relation to an integrable deformation of CFT . 
%So, if the AdS/CFT generically has such an interpretation between cutoff in gravity side and a deformation of gauge theory, our calculation might have some implication of a deformation of five dimensional super Yang-Mills theory. 
%} 
\begin{align}
&X_{1}(\tau, \sigma = \delta) = \frac{L_{q}}{2} + \epsilon \alpha(\tau ; \Delta \tau), \label{bc1}\\
&X_{1}(\tau, \sigma = \pi-\delta) = -\frac{L_{q}}{2} - \epsilon \alpha(\tau ; \Delta \tau), \label{bc2}
\end{align}
where $\alpha(\tau; \Delta \tau)$ is defined by
\begin{align}
\alpha(\tau; \Delta \tau) = 
\begin{cases}
\text{exp}\left[ 2 \left( 4-\frac{\Delta \tau}{\tau} - \frac{\Delta \tau}{\Delta \tau - \tau} \right) \right], ~ 0 < \tau < \Delta\tau & \\
0, \qquad\qquad\qquad\qquad\qquad\quad \text{otherwise} &
\end{cases}
\end{align}
In our numerical simulation, we introduce a small cutoff $\delta$ near string endpoints 
and set our numerical domain in $\delta\leq \sigma \leq \pi-\delta$.
As long as we take a small enough cutoff $\delta$, it may not matter to our results as shown in appendix \ref{sec:error}.\footnote{
This cutoff might be understood as an integrable deformation of dual field theories, see
\cite{Smirnov:2016lqw,Cavaglia:2016oda,McGough:2016lol,Kraus:2018xrn,Chakraborty:2018aji}.}
There are two parameters $\epsilon$ and $\Delta\tau$, which are the amplitude and the time scale of the quench. 
One can check that $\alpha(\tau; \Delta \tau)$ is $C^{\infty}$ in all $\tau$ and has a compact support in the region $0 \leq \tau \leq \Delta \tau$. 

\begin{figure}[tb]
\includegraphics[width=70mm, trim=100 50 50 20]{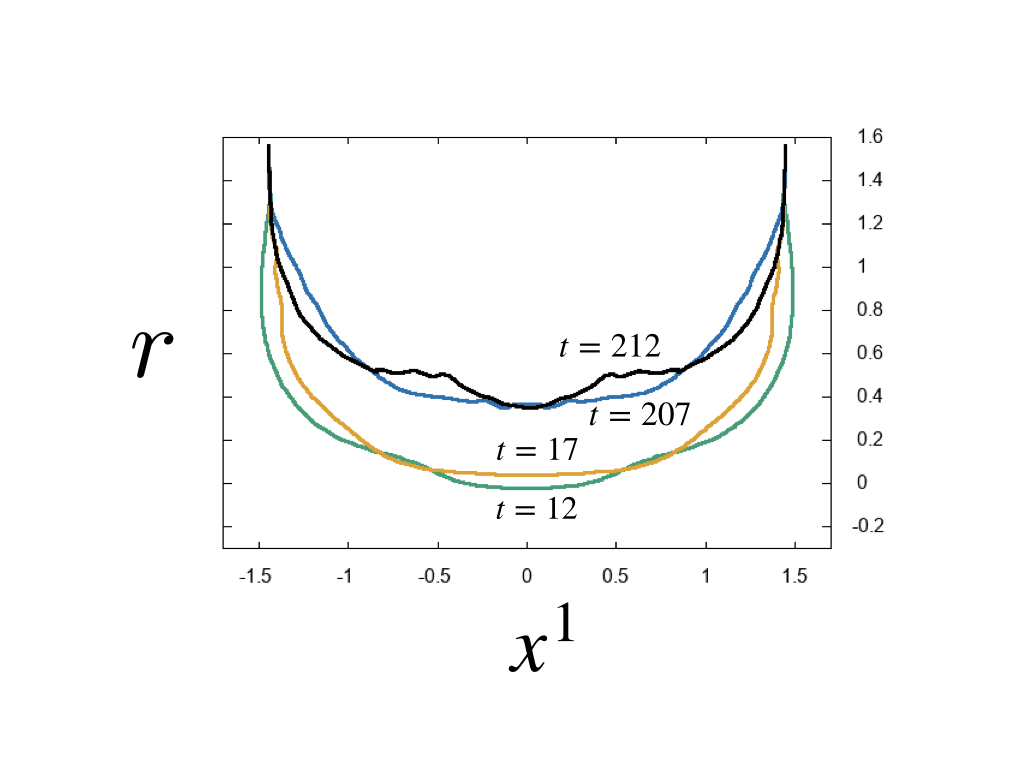}
\caption{\label{fig:snapshot} Snapshot of dynamical strings. }
\end{figure}

Boundary conditions for the other variables at 
time-like boundaries are $X_{2} = X_{3} = 0$ and $R = R_\textrm{ini}$, where
$R_\textrm{ini}\equiv R(\tau=0,\sigma=\delta)$ is the initial value of the $R$ at the boundary.
Because of the trivial boundary conditions of $X_2$ and $X_3$, they are identically zero throughout time evolution.
The boundary value of $T$ is determined by constraints $C_{u}=C_{v} = 0$. 
(See appendix \ref{sec:timeev} for the detail.)
With these boundary conditions, the string motion is $\mathbb{Z}_{2}$-symmetric under $x^{1}\to -x^{1}$. 
Fig.~\ref{fig:snapshot} shows the string configuration
in the $x^{1}$-$r$ plane for $\epsilon = 0.4$ and $\Delta \tau = 4$. 
%(\textbf{$\Delta v=2$で計算してるから、$\Delta \tau=4$でない?要確認。})
In this figure, we took a time slice of bulk coordinate as $t= T(u, v)$. 
Fig.~\ref{fig:snapshot} contains snapshots of string at early and late times. 
At early times, string profiles seem smooth. 
On the other hand, at late times small spatial deformations are observed. 
We also monitored violation of constraints~(\ref{conu}, \ref{conv}) 
and found that they are sufficientlly small. (See appendix \ref{sec:error}).

% Fig.  snapshot of string motions, and the plot of r_{center}

\begin{figure}[tb]
\includegraphics[width=65mm, trim=100 50 50 20]{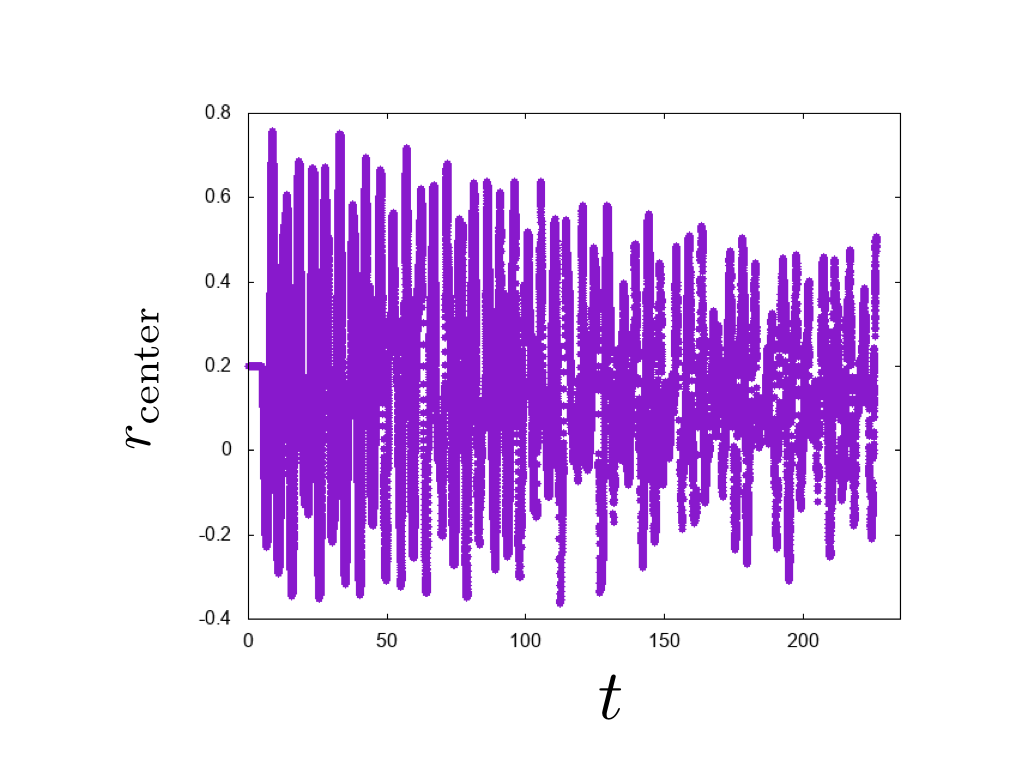}
\caption{\label{fig:rcenter} The trajectory of the tip of the dynamical string for $r_{\text{center}} = 0.2,\, \epsilon = 0.4$, and $\Delta\tau = 4$. }
\end{figure}

To observe its chaos,  
we focus on the tip of the string, $r_{\text{center}} = R(\tau, \sigma=\pi/2)$, since
the string motion is completely $\mathbb{Z}_{2}$-symmetric due to our boundary condition and the 
quench.
Fig.~\ref{fig:rcenter} shows time dependence of $r_{\text{center}}(t)$.
Chaos means a sensitivity to the change of the initial conditions.
To explore the sensitivity of the string motion, we consider a linear perturbation: $T\to T+\delta T, R\to R+\delta R$ and $\vec{X} \to \vec{X}+\delta\vec{X}$. 
We numerically solve the linear evolution equations for $(\delta T, \delta R, \delta\vec{X})$ on the time dependent background $(T(u,v), R(u,v), X_1(u,v))$. 
Initial conditions are $0$ for all variables and boundary conditions are $\delta X_{1}(\tau,\sigma =\delta,\pi-\delta) = \pm \alpha(\tau; \Delta\tau=4)$, and $(\delta R, \delta X_{2}, \delta X_{3})|_{\sigma=0,\pi} = 0$. 
The boundary conditions for $\delta T$ at $\sigma = \delta,\pi-\delta$ are again determined by linearized constraints $\delta h_{uu} = \delta h_{vv} =0$.

%Fig. time evolution of delta r_{center} and its exponential grouwth
\begin{figure}[tb]
\includegraphics[width=65mm, trim=160 50 30 20]{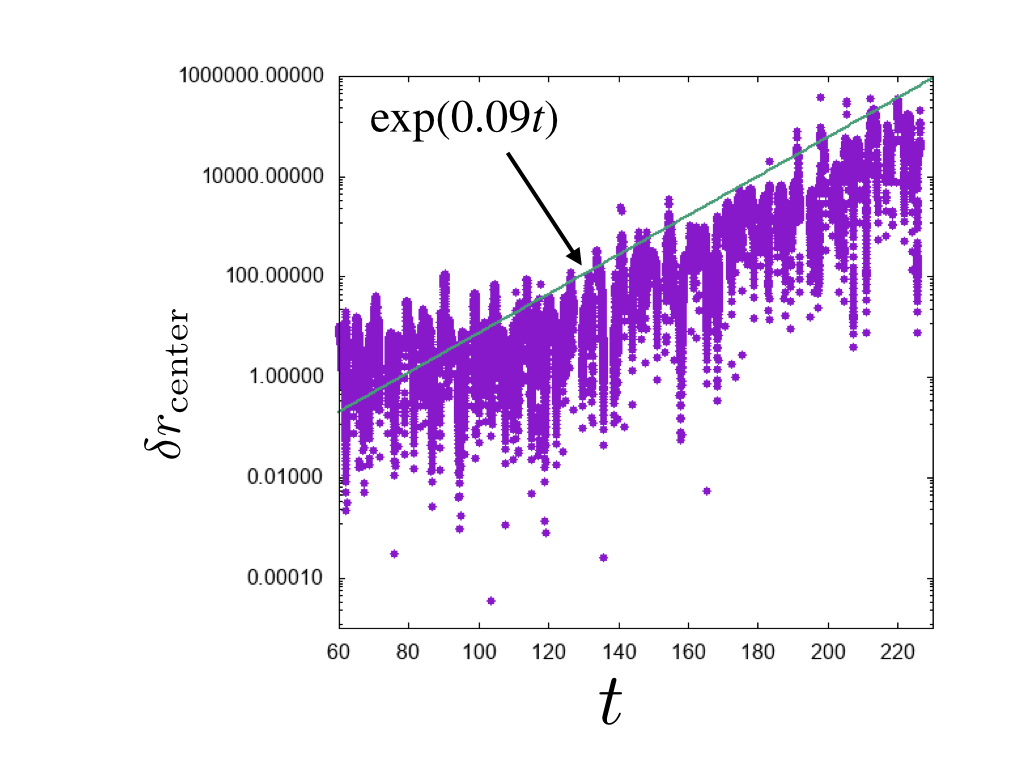}
\caption{\label{fig:chaos} Sensitivity of $r_{\text{center}}$, the tip of string, to the initial perturbation. The figure shows that $\delta r_{\text{center}}$ exponentially grows and the Lyapunov exponent can be read off from the coefficient of $t$: $\lambda_{L}\simeq 0.09$. }
\end{figure}

% Fig. phase diagram: horizontal axis Lq, vertical axis epsilon
\begin{figure}[tb]
\includegraphics[width=70mm, trim=100 60 80 40]{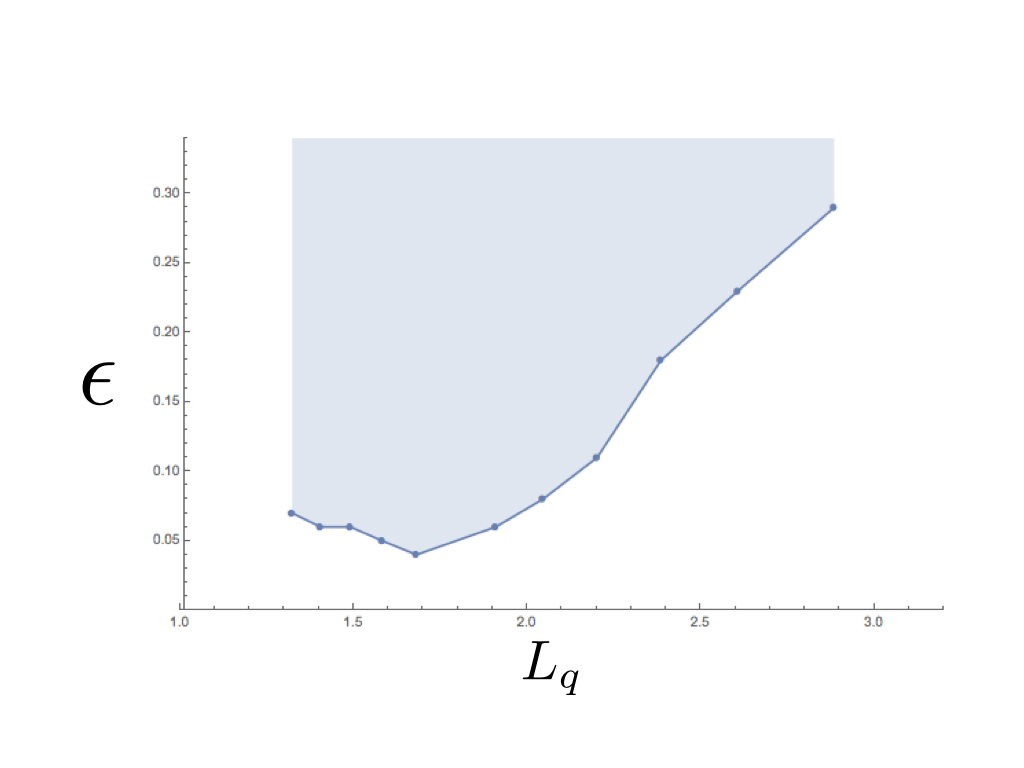}
\caption{\label{fig:phasediagram} Phase diagram of the chaos of the QCD string. For each fixed interquark distance $L_{q}$, we numerically solve the string motion with different $\epsilon$, the amplitude of the pulse force (quench) acting on the quarks. In the shaded region we find that the motion is potentially chaotic, and below that chaos does not appear. This implies that for larger $L_{q}$ the motion is less chaotic. We do not plot the region $L_{q} < 1.2$, since for a small $L_{q}$ the gauge theory becomes five-dimensional, which is not QCD-like.}
\end{figure}

The results of time evolution of the linear perturbations are as follows. 
Fig.~\ref{fig:chaos} shows a time evolution of $\delta r_{\text{center}}(t)$, the tip of the string
(located at $X_1=0$ due to the $\mathbb{Z}_{2}$ symmetry), for $\epsilon= 0.4$. 
The horizontal axis is the bulk time coordinate $t = T(\tau, \sigma = \pi/2)$. 
We can find an exponential growth of the initial perturbation, which implies chaos. 
Fitting the amplitude, we obtain the positive Lyapunov exponent as $\lambda_{L} \simeq 0.09$ . 

In the numerical simulations, 
we also observed that for small enough $\epsilon$ chaos does not occur, which means that 
there may be a chaos threshold bound of $\epsilon$ for each initial configuration given by $L_q$
(corresponding to the initial condition for $r_{\text{center}}$).  
We investigate the bound by running the simulation for different values of the 
parameters: quark separation $L_{q}$ and $\epsilon$.
% which correspond to the initial condition for $r_{\text{center}}$ and the amplitude of quench, respectively. 
Our final ``phase diagram'' of the chaos of the QCD string is shown in Fig.~\ref{fig:phasediagram}.
%It implies that in confining phase the string motion is less chaotic and near transition point from confining to Coulombic phase chaos seems easy to occur. 
%One comment on Fig  epsilon
%One comment on the phase diagram of chaos, Fig. \ref{fig:phasediagram}. 
For each fixed quark-antiquark separation $L_{q}$, we numerically solve the full dynamical motion of string with different amplitude of quench $\epsilon$ and study whether the motion is chaotic. 
Below the solid line in Fig.~\ref{fig:phasediagram} the motion is regular, while above the line chaos appears.
%\footnote{ 
%Due to our numerical limitation, for too large $\epsilon$ the numerical calculation breaks down before we can read off the Lyapunov exponent. 
%Thus, we observe chaos right above the solid line, and we conclude that the dynamical motion of string is potentially chaotic in the shaded region in Fig.~\ref{fig:phasediagram}. 
%}

The phase diagram (Fig.~\ref{fig:phasediagram}) shows the following two important behavior:
First, there exists a lower bound for the magnitude of the boundary pulse force $\epsilon$,
for the chaos to occur. Second, the bound is a function of the interquark distance $L_q$, and
it grows as $L_q$ grows. This in particular means that long strings are less chaotic.
The shape of the bound described in Fig.~\ref{fig:phasediagram} appears to be consistent with what 
we obtained in the rectangular string toy model in the previous section, Fig.~\ref{fig:Echaos}.
We shall investigate more on this behavior of the chaos bound in the next section, by using a physical model, to locate the origin of the chaos.

%Fig. chaos of interquark force
\begin{figure}[tb]
\includegraphics[width=65mm, trim=100 50 50 0]{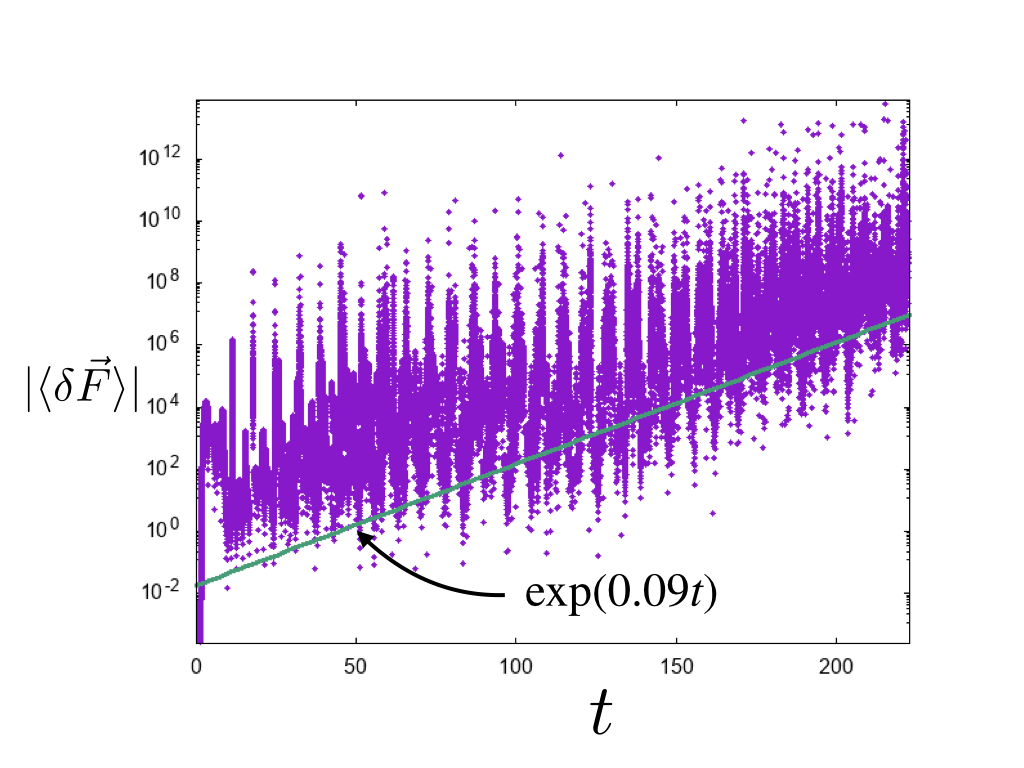}
\caption{\label{fig:force} Sensitivity of the interquark force to the initial perturbation. 
We here again read off the positive Lyapunov exponent of the chaos, and find that it is consistent with that of $\delta r_{\text{center}}$. This holographically implies that the force acting on quarks in the gauge theory is chaotic. }
\end{figure}

Finally, let us provide a prediction about a QCD quantity.
We can also observe positive Lyapunov exponents for an observable of the gauge theory. 
When the string endpoint does not move, the AdS/CFT tells us that the force acting on the quark and the antiquark in the gauge theory is given by
\begin{align}
\langle \vec{F}(t) \rangle = \frac{\lambda}{72\pi} \partial_{r}^{4}\vec{X}|_{r=\pi/2}, 
\label{eq:force}
\end{align}
where $\lambda$ is the 't Hooft coupling appearing as an overall coefficient in \eqref{eq:nullaction}. 
The derivation of \eqref{eq:force} is given in App.~\ref{sec:force}. 
Fig. \ref{fig:force} shows the time evolution of $\langle \delta \vec{F}(t) \rangle$ and it implies the sensitivity of the interquark force to an initial perturbation. 
$\langle \delta \vec{F}(t) \rangle$ grows exponentially and its Lyapunov exponent is consistent with that of $\delta r_{\text{center}}(t)$. 
We find chaos of the interquark force via the AdS/CFT: the force in large $N_{c}$ pure Yang-Mills theory is generically sensitive to initial perturbations. 
%(\textbf{$\delta F$は数値的にどうやって評価したんでしたっけ？簡単にかけるなら、appendix.Bかここらへんに書いてください。})

%%%%%%%%%%%%%%%%%%%%%%%%%%%%%%%%%%%%%%%%%%%%%%%%%%

\section{\label{sec:model}Possible origin of the chaos}

We found in the numerical simulation that the NG string in the confining geometry shows chaos,
when the energy of the string exceeds some lower bound which is a function of the 
interquark distance $L_q$. In this section we argue why this behavior appears,
based on a simple argument. 

\begin{figure}[tb]
\includegraphics[width=60mm, trim=100 60 80 40]{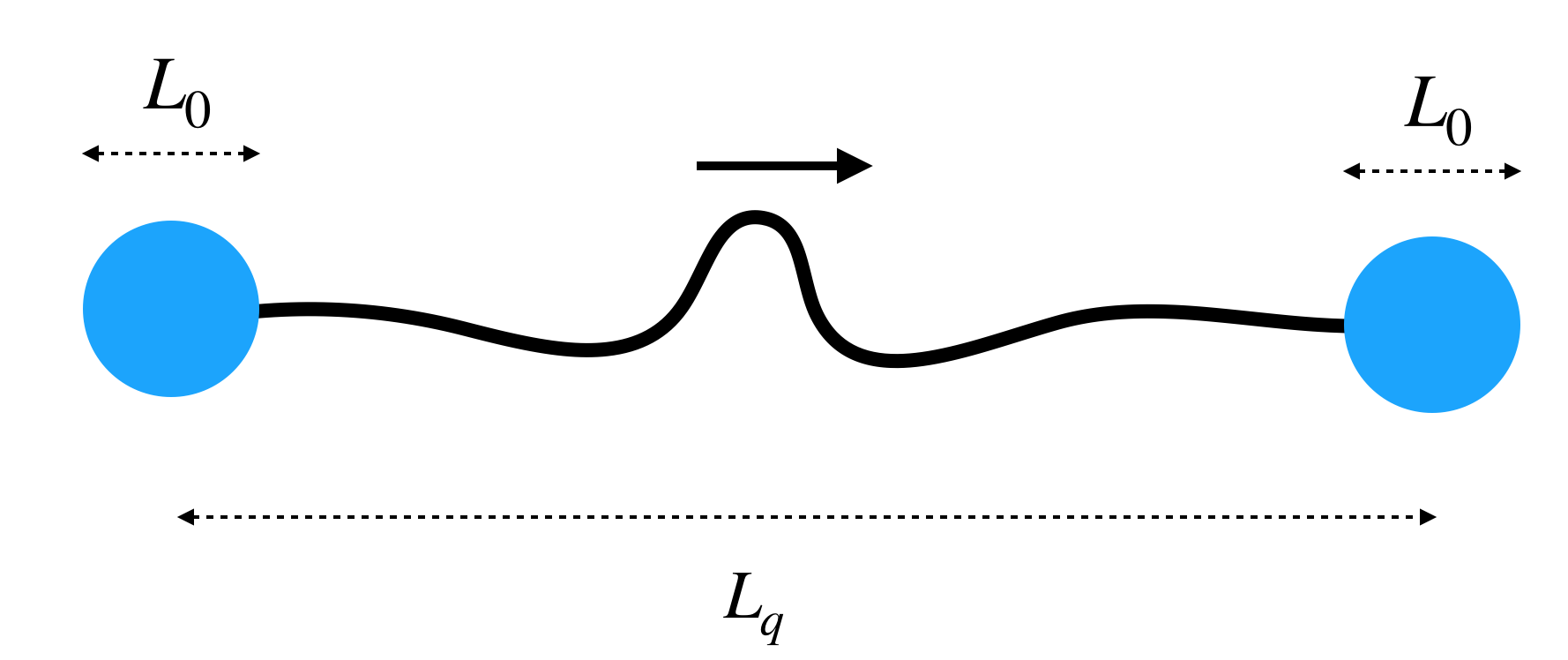}
\vspace*{5mm}
\caption{\label{fig:model-chaos} A string model with boundary quarks of the size of the QCD scale $\sim L_0$. }
\end{figure}

First of all, in the rectangular string model in Sec.~\ref{sec:rect}, the chaos shows up
as a result of the different oscillation modes at the bottom of the string.
On the other hand, it is known that straight string is integrable. This leads us 
to suspect that the
origin of the chaos should be at the boundaries of the string. 
Let us consider a string model shown in Fig.~\ref{fig:model-chaos}:
%. It models a model made of a 
%QCD string:
Two quarks are connected by a long straight string in the three-dimensional space. 
On the string any wave can propagate,
and the motion is integrable. The wave will hit the boundary which is a quark. The boundary 
is not a point, but a region of the QCD scale. Since the string propagation part is integrable,
any chaos, if exists, should originate in the boundary regions. We naively assume that when the
magnitude of a wave hitting the boundary region exceeds some threshold value $\epsilon_0$ 
the chaos emerges. 
The wave amplitude will decay while it propagates, and so, the system with a 
larger interquark distance $L_q$ is expected to be less chaotic.

% Fig. fitted phase diagram

To quantify this physical model, we solve a motion of the wave propagating on the straight string.
If the NG string sits at the bottom of the geometry, the fluctuation of the string obeys the
wave equation
\begin{align}
\left[\partial_t^2-\partial_x^2 + M_{\rm KK}^2\right] U(t,x)=0 \, .
\end{align}
The mass can be obtained by the analysis of the fluctuation of the straight string.
A typical solution with a momentum $k_0$ larger than the mass scale, $k_0 \gg M_{\rm KK}$,
is
\begin{align}
U(t,x) = \int dk \, f(k) \exp\left[
it \sqrt{k^2+M_{\rm KK}^2} - ik x
\right]
\end{align}
where $f(k)$ is centered at $k=k_0$. Expanding this for small $M_{\rm KK}/k_0$, we obtain
\begin{align}
U \propto \cos\left[\frac{M_{\rm KK}^2}{2k_0}t\right] \sim 1-\frac{M_{\rm KK}^4}{8k_0^2}t^2 + \cdots \, .
\end{align}
which means that the amplitude of the fluctuation decays along the propagation on the string.
The timescale for the fluctuation to reach the other side of the string is estimated as
$t \sim L-L_0$
where $L_0/2$ is the size of the boundary region which is expected to be the QCD scale.
The typical momentum $k_0$ is estimated as $k_0 \sim \pi/(2\Delta \tau)$ for the initial kick
in our numerical simulation. Using these, the lower bound for the chaos is given by
\begin{align}
\epsilon \geq \epsilon_{0} \left( 1+ \frac{ M_{\rm KK}^4(\Delta\tau)^2}{2\pi^{2}}
 (L_{q}-L_{0})^{2} \right).
\label{eq:fit}
\end{align}
This expression shows that a larger $L_q$ makes the chaos diminished.

\begin{figure}[tb]
\includegraphics[width=70mm, trim=100 60 80 40]{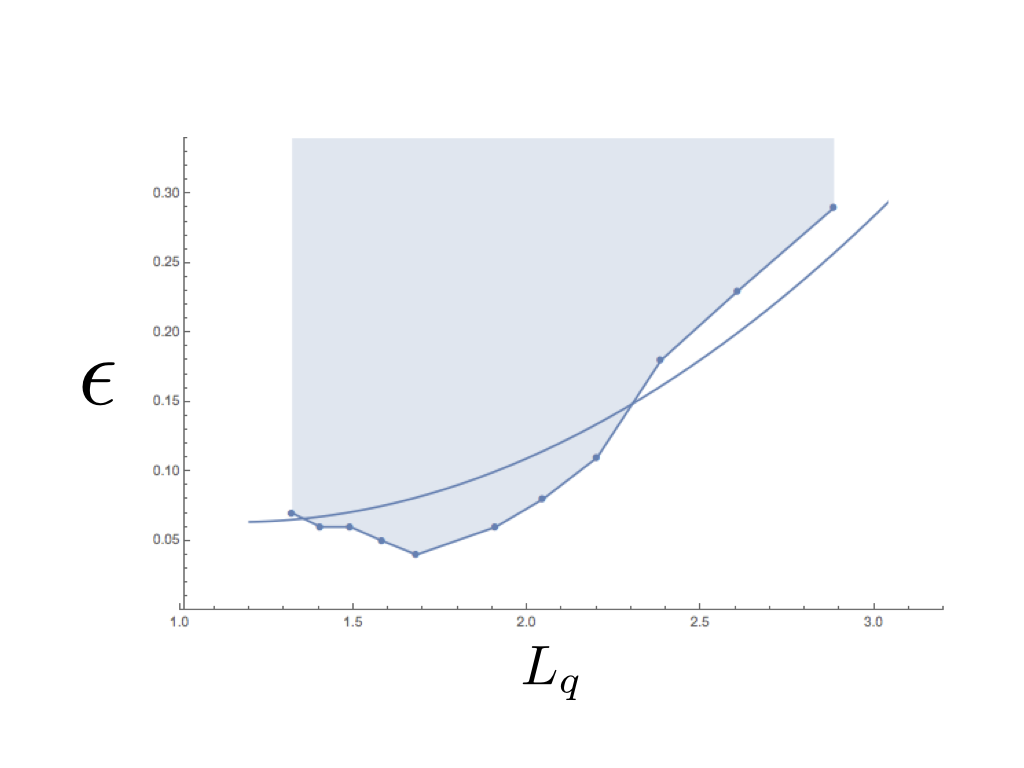}
\caption{\label{fig:phasefit} Phase diagram fitted by the quadratic function (\ref{eq:fit}). }
\end{figure}

By this analytic expression \eqref{eq:fit}, we can fit our numerical lower bound of the chaos, Fig.~\ref{fig:phasediagram}. Our numerical simulation uses $M_{\text{KK}}=3/2$, $\Delta\tau=4$. 
We find that choosing 
$\epsilon_{0}=0.064$ and $L_{0} = 1.2$ fits the numerically obtained bound qualitatively, 
see Fig.~\ref{fig:phasefit}. The obtained value, $L_0/2 \sim 0.6$, roughly coincides with $1/M_{\rm KK}$ which is the QCD scale of the model.

% Fig. fitted phase diagram

From this argument, we find that a physical picture consistent with the results of the numerical simulation is a quark model in which an integrable string connect two boundaries whose size is
of the QCD scale, and the boundary region produces chaos if the input wave exceed a certain
threshold amplitude. The chaos originates in the boundaries of the QCD string, the constituent 
quarks.

%%%%%%%%%%%%%%%%%%%%%%%%%%%%%%%%%%%%%%%%%%%%%%%%%%

\section{\label{sec:summary}Summary and Discussion}

In this paper, we studied chaos and time evolution of interquark force by using the AdS/CFT correspondence. 
%To study the time evolution of Wilson loop, 
We performed a full nonlinear numerical simulation of the dynamics of a NG string in the confining geometry in the gravity side. 
The AdS/CFT translates the chaos of the NG string to the chaos of the interquark force. 
We found that the interquark force in large-$N_{c}$ four-dimensional pure Yang-Mills theory is generically sensitive to initial perturbations, and it is actually chaotic.

%In Sec. \ref{sec:static}, we first reviewed confining geometry in gravity side. 
%Solving the static equation of motion of an open string hanging on the D4 boundary, we confirmed that it holographically realizes confining potential in the dual gauge theory.
%In Sec. \ref{sec:chaos}, we further developed 
Our numerical calculation of the string in the D4-soliton background
enabled us to analyze the full dynamical motion in details, and the Lyapunov exponent
was obtained.
Using the AdS/CFT dictionary, we further obtained the Lyapunov exponent of the interquark force.
%In addition, we can find the Lyapunov exponent of interquark force by relating it with the dynamics of an open string via holography, Eq. (\ref{eq:force}). 
%The point is that we directly computed the time evolution of interquark force, which is 
Normally, time-dependence of gauge-invariant non-perturbative observables of QCD is quite difficult
to compute, 
% a quite difficult problem in perturbative gauge theory or in a simulation of lattice QCD. 
thus, our results provide a theoretical prediction: 
the dynamics of the non-perturbative Yang-Mills gauge theory may be generically chaotic. 

Our numerical simulations have two adjustable parameters: the interquark distance $L_q$ and
the strength $\epsilon$ of the impulse force on the quarks to make them start moving.
By area-bombing the parameter space, we obtained a phase diagram of the chaos, Fig.~\ref{fig:phasediagram}. It exhibits a unique picture: there exists a lower bound
of $\epsilon$ for the chaos to occur, and the bound grows as $L_q$. 
This feature can be understood if the chaos originates in the constituent quark sectors
(which are the boundaries of the QCD string),
as provided in Sec.~\ref{sec:model} with a simple model. 

We provided a prediction of the Lyapunov exponent for the interquark forces. We hope we can 
confirm the exponent by some other direct calculations of QCD. 
%EMT distribution
%In this paper, we considered time evolution of Wilson loop and force acting on quarks. 
Recently, the gradient flow techniques have been applied to lattice QCD simulations and the energy-momentum tensor on the lattice was defined through a flow equation \cite{Suzuki:2013gza}. 
By using these techniques, the three dimensional distribution of energy-momentum stress tensor in $SU(3)$ gauge theory is non-perturbatively computed \cite{Yanagihara:2018qqg}. 
However, the lattice QCD analyses are still only for static observables, and it is difficult
to follow the time dependence. Nevertheless, it would be beneficial to compare the
structure of the lattice QCD string with the holographic QCD string and find some difference,
to locate possible origin of chaos qualitatively.
%Therefore, reproducing this results holographically and studying the time evolution of full energy-momentum tensor may be a good future direction. 

Our study focused on light modes of the large $N_c$ QCD, which are mesons and glueballs, while
heavy nonlocal excitations exist: baryons and nuclear resonances. 
It would be important to quantify chaos of large $N_c$ baryons and nuclei
and compare them with that of mesons and the QCD strings to find any difference in origin.
Again, holography can help analyzing the chaos of the single or multiple baryon(s). They are known to be dual
to D-branes called baryon vertices \cite{Witten:1998xy} in the gravity side, so the motion of
the baryons are well-approximated by a dimensionally reduced Yang-Mills theories \cite{Hashimoto:2008jq}.
Based on the classical 1-dimensional Yang-Mills analyses \cite{Matinyan:1981dj,Matinyan:1986nw} 
and on their D0-brane interpretation \cite{Gur-Ari:2015rcq,Asano:2015eha}, or more detailed ADHM-like
matrix model formulation \cite{Hashimoto:2010je} and its quantum states \cite{Hashimoto:2019wmg}, 
it is possible to quantify the chaos of baryons.
Since it is known that nuclear resonances follow quantum chaos \cite{Haq:1982soh},
finding out random matrix-like behavior from the classical holographic baryons would be interesting.

The chaos in the gravity side has been studied in the context of black hole horizons and the infinite redshift. The universal chaos bound discovered in \cite{Maldacena:2015waa} is $\lambda \leq 2\pi T$
for large $N_c$ system with a finite temperature $T$,
and it is proven that all observables in the large $N_c$ limit should obey this chaos bound
for the quantum Lyapunov exponent defined by the out-of-time ordered correlators.
Our case is at zero temperature, so, if we naively apply the chaos bound to the zero-temperature 
large $N_c$ QCD, any chaos is not allowed. This appears to contradict with our finding that
the interquark force has a nonzero Lyapunov exponent and thus is chaotic --- apparently there
should be a loophole. The point is that
the bound in \cite{Maldacena:2015waa} was for local operators, while our observables are non-local, so
the bound does not apply naively. Since non-Abelian gauge theories are always accompanied 
by non-local observables, it would be interesting to study how the quantum Lyapunov exponent of
those non-local observables in generic gauge theories is theoretically observed, and how they play 
a role in determining the spectral/dynamical aspects of generic gauge theories.

\begin{acknowledgments}
We would like to thank Tadakatsu Sakai, Motoi Tachibana, and Ryosuke Yanagihara for variable discussions. 
The work of K.H. was supported in part by JSPS KAKENHI Grants No.~JP15H03658, No.~JP15K13483, and No.~JP17H06462.

\end{acknowledgments}

\appendix

\section{Numerical details}
\label{sec:num}

\subsection{Initial data}
\label{sec:initial}

As the initial data, we use the static string configuration.  
Here, we explain how to express the static solution in the double null coordinate $(u,v)$.
Introducing $\tau=u+v$ and $\sigma=u-v$, 
we assume that the static solution is written as 
$T=\tau$, $\vec{X}=(X(\sigma),0,0)$ and $R=R(\sigma)$.
In this assumption, Eq.(\ref{Tuv}) is automatically satisfied.
Integrating Eq.(\ref{Xuv}) by $\sigma$, we obtain
\begin{equation}
 X'=\frac{\cos^3 R}{\cos^3 r_{\rm center}}\ ,
\label{Xprime}
\end{equation}
where $r_{\rm center}$ is the integration constant and ${}'\equiv d/d\sigma$.
Substituting above expression into constraints~(\ref{conu}) and (\ref{conv}), we have 
\begin{multline}
 R'{}^2=\frac{1}{4}(1+\cos^2R+\cos^4R)\left(1-\frac{\cos^6 R}{\cos^6 r_{\rm center}}\right)\ .
\label{rprime}
\end{multline}
At $R=r_{\rm center}$, we have $R'(\sigma)=0$. Thus, 
$R=r_{\rm center}$ corresponds to the position of the tip of the hanging string.
Note that this equation is regular at $R=\pi/2$ and well-behaved near the AdS boundary.
On the other hand, near the tip of the hanging string, $R'\sim \sqrt{r_{\rm center}-R}$. 
This is not a suitable form for the numerical integration around $R=r_{\rm center}$.
From Eq.(\ref{ruv}), we can obtain the other equation for $R(\sigma)$ as
\begin{multline}
 R'' = -\frac{3}{2}R'{}^2 \tan R \bigg[ 1+ \frac{3\cos^{2}R (1+2\cos^{2}R)}{1+\cos^{2}R + \cos^{4}R} \bigg] \\
+\frac{3}{8}\tan R (1+\cos^{2}R + \cos^{4}R) (1 + \frac{\cos^6 R}{\cos^6 r_{\rm center}}).
\label{rprime2}
\end{multline}
This can also be derived by differenciating Eq.(\ref{rprime}) by $\sigma$.
Contrary to Eq.(\ref{rprime}), above equation is singular at $R=\pi/2$ but regular at $R=r_{\rm center}$.
Therefore, in our numerical construction of the initial data, 
we integrate Eq.(\ref{rprime2}) from $R=r_{\rm center}$ to $R=(\pi/2+r_{\rm center})/2$. 
We then switch the equation to Eq.(\ref{rprime}) and continue the integration from $R=(\pi/2+r_{\rm center})/2$ to $R=\pi/2$.
Once we have the numerical solution of $R(\sigma)$, we also obtain $X(\sigma)$ integrating Eq.(\ref{Xprime}).
As the result, we have the right half of the static string in Fig.\ref{potential}.
We reparametrize the worldsheet coordinates as 
$\tau\to c \tau$ and $\sigma\to -c \sigma + c'$ ($c$ and $c'$ are constants) 
so that $R|_{\sigma=0}=\pi/2$ and $R|_{\sigma=\pi/2}=r_{\rm center}$ are satisfied.
The left half of the static string can be easily generated by the $\mathbb{Z}_{2}$-symmetry: 
$R(\sigma)=R(\pi-\sigma)$ and $X(\sigma)=-X(\pi-\sigma)$.
Then, time-like boundaries of the worldsheet are located at $\sigma=0,\pi$.

\subsection{Time evolution}
\label{sec:timeev}

The original form of evolution equations~(\ref{Tuv}-\ref{Xuv}) is numerically unstable.
To stabilize time evolution, 
we eliminate $T_{,u}$ and $T_{,v}$ from Eqs.(\ref{Tuv}-\ref{Xuv}) 
using the constraint equations~(\ref{conu}) and (\ref{conv}).
Resultant equations are written in the form of
\begin{equation}
 \Phi_{,uv}=\bm{F}(\hat{\Phi}_{,u},\hat{\Phi}_{,v},\hat{\Phi})\ .
\label{Phieq}
\end{equation}
where $\Phi=(T,R,\vec{X})$, $\hat{\Phi}=(R,\vec{X})$ and 
$\bm{F}$ is a non-linear function of its arguments.

\begin{figure}[tb]
\includegraphics[scale=0.5]{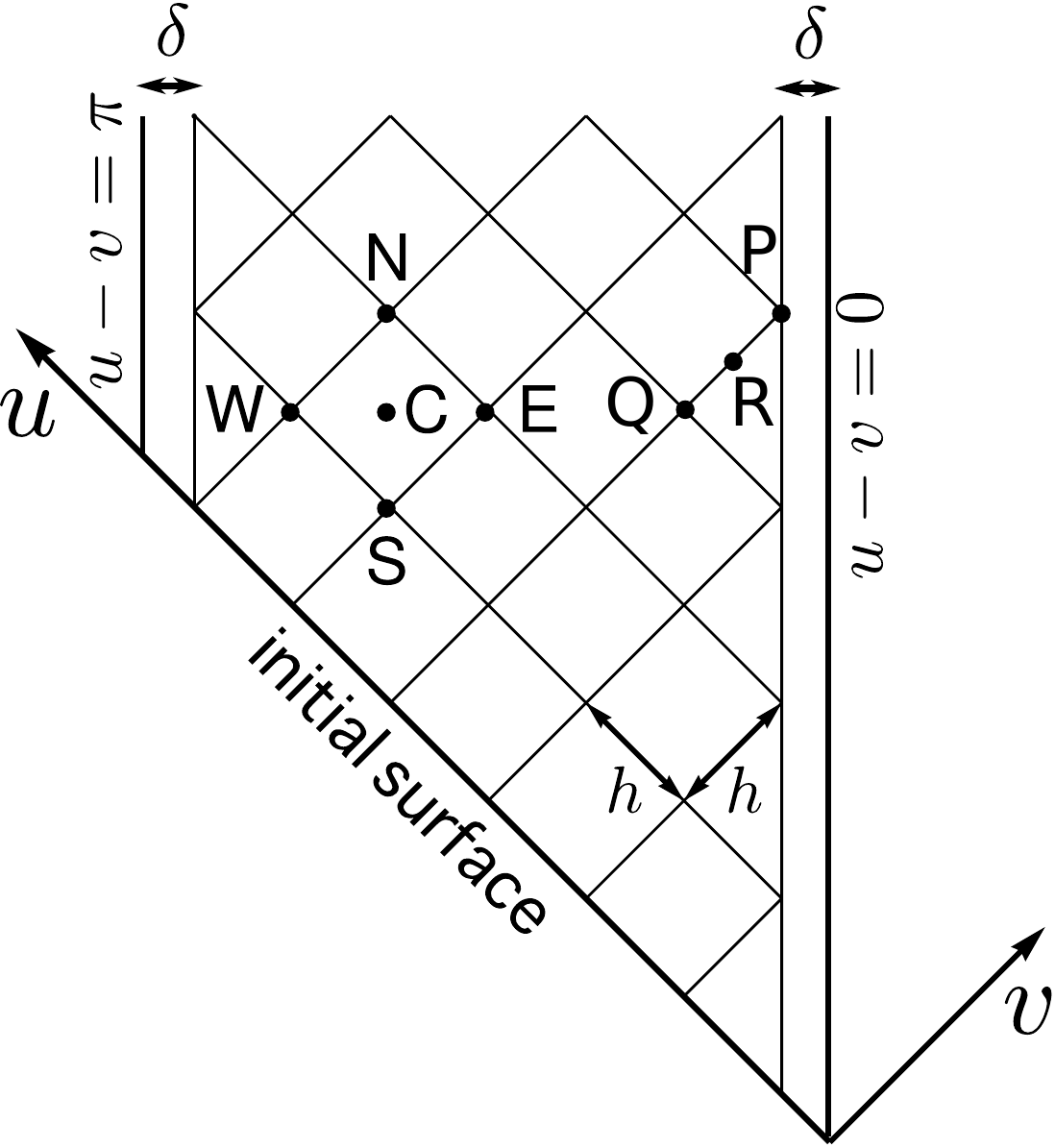}
\caption{\label{fig:grid} 
Grid points on the worldsheet for numerical calculations.
A small cutoff $\delta$ is introduced near time-like boundaries.
}
\end{figure}

We take uniform grid along $u$ and $v$ as in Fig.\ref{fig:grid}.
The grid points are explicitly written as $v=jh$ and $u=(i+j)h + \delta$ ($i=0,1,2,\cdots,N$, $j=0,1,2,\cdots$.)
where $h=(\pi-2\delta)/N$ is the mesh size and $N$ is the number of grid points along the $u$-direction.
Our numerical domain is in $\delta\leq u-v \leq \pi-\delta$ and $v\geq 0$.
We introduced a small cutoff $\delta$ near the time-like boundaries of the worldsheet.
If we set exactly $\delta=0$, the numerical simulation immediately breaks down and we cannot even see regular time evolutions. 
%We found that a finite cutoff is needed to stabilize numerical time evolution of the string in D4-soliton geometry.
%although we can take $\delta=0$ for asymptotically AdS cases~\cite{Ishii:2015wua,Ishii:2015qmj,Hashimoto:2018fkb}, 

Let us focus on points N, E, W, S and C in Fig.\ref{fig:grid}. We can evaluate $\Phi$ and its derivatives at the point C 
with second-order accuracy in $h$ as
$\Phi_{,uv}|_\textrm{C}=(\Phi_\textrm{N}-\Phi_\textrm{E}-\Phi_\textrm{W}+\Phi_\textrm{S})/h^2$, 
$\Phi_{,u}|_\textrm{C}=(\Phi_\textrm{N}-\Phi_\textrm{E}+\Phi_\textrm{W}-\Phi_\textrm{S})/(2h)$, 
$\Phi_{,v}|_\textrm{C}=(\Phi_\textrm{N}+\Phi_\textrm{E}-\Phi_\textrm{W}-\Phi_\textrm{S})/(2h)$ and 
$\Phi|_\textrm{C}=(\Phi_\textrm{E}+\Phi_\textrm{W})/2$, 
where $\Phi_{\textrm{N,E,W,S}}$ denote numerical values of $\Phi$ at points N,E,W,S.
Substituting them into Eq.(\ref{Phieq}), we obtain the discretized version of the evolution equation.
The equation determines $\Phi_\textrm{N}$ from $\Phi_\textrm{E,W,S}$.
We use the Newton-Raphson method for solving the equation.

Once we have the initial data at $v=0$ and boundary data at $u-v=\delta, \pi-\delta$, 
we can determine the solution in our numerical domain by solving the discretized equation. 
As the initial data, we use the static string obtained in Sec.\ref{sec:initial}.
(So, the constraint~(\ref{conu}) is satisfied at $v=0$.)
At boundaries $u-v=\delta, \pi-\delta$, we do not change $R$ from its initial value: 
$R(\tau,\sigma=\delta)=R(\tau=0,\sigma=\delta)$ and $R(\tau,\sigma=\pi-\delta)=R(\tau=0,\sigma=\pi-\delta)$.
We impose the Dirichlet conditions for $X_1$ as in Eqs.(\ref{bc1}) and (\ref{bc2}).
To determine the boundary value of $T$, we consider points P, Q, and R in Fig.\ref{fig:grid}.
We can evaluate $\Phi$ and its $v$-derivatives at the point R
as $\Phi_{,v}|_\textrm{R}=(\Phi_\textrm{P}-\Phi_\textrm{Q})/h$ and 
$\Phi|_\textrm{R}=(\Phi_\textrm{P}+\Phi_\textrm{Q})/2$. 
Substituting them into the constraint equation~(\ref{conv}), we have the equation for $T_\textrm{P}$.
By the similar way, using the other constraint~(\ref{conu}), we obtain the left boundary value of $T$.

Substituting $\Phi \to \Phi + \delta \Phi$ into Eq.(\ref{Phieq}) and taking first order in $\delta\Phi$, 
we obtain the linear partial differential equation for $\delta \Phi=(\delta T, \delta R, \delta X)$.
We also solve the evolution of the linear perturbation numerically. 
Its numerical procedure is completely parallel to that for the background.

\subsection{Error analysis}
\label{sec:error}

\begin{figure}[tb]
  \centering
\subfigure[]
 {\includegraphics[scale=0.35]{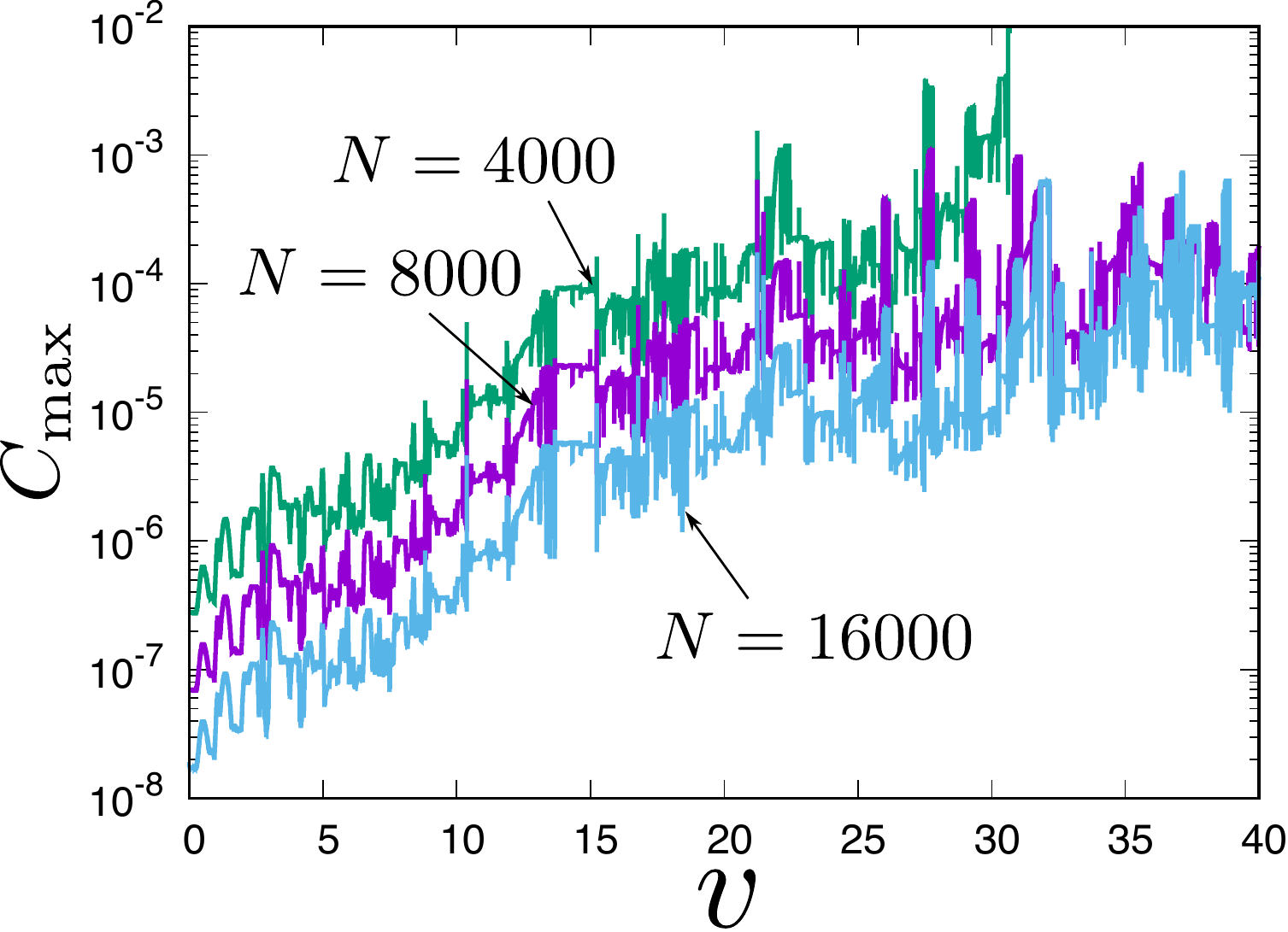}\label{const_Ndep}
  }
\subfigure[]
 {\includegraphics[scale=0.4]{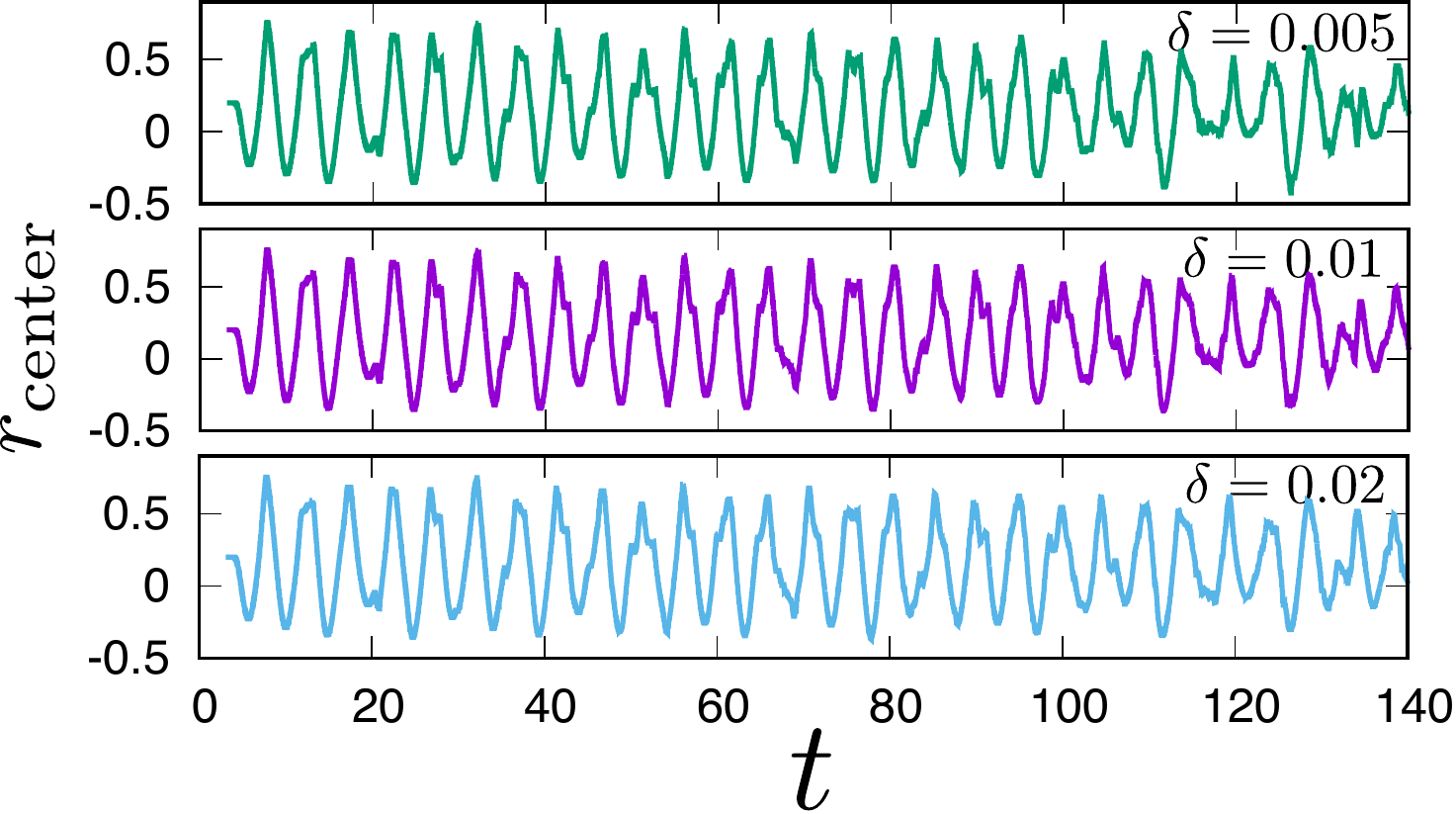}\label{cutoff_dep}
  }
 \caption{(a)Constraint violation $C\textrm{max}$ for $N=4000,8000,16000$. 
(b)Time dependence of the tip of the string $r_\textrm{center}(t)$ for several values of cutoff $\delta=0.005,0.01,0.02$.
 }
 \label{errorana}
\end{figure}

As the measure of the numerical error, 
we monitior the violation of the constraints~(\ref{conu}) and (\ref{conv}).
We introduce the normalized constraint as
\begin{equation}
 C(u,v)=\frac{|C_u|+|C_v|}{1+\mathcal{N}_u+\mathcal{N}_v}\ ,\label{Cuv}
\end{equation}
where $\mathcal{N}_u$ and $\mathcal{N}_v$ are ``scales'' of constraints:
\begin{align}
 \mathcal{N}_u&=T_{,u}^2 +\vec{X}_{,u}^2 +\frac{4R_{,u}^2}{1+\cos^{2}R + \cos^{4}R}\ ,\\
 \mathcal{N}_v&=T_{,v}^2 +\vec{X}_{,v}^2 +\frac{4R_{,v}^2}{1+\cos^{2}R + \cos^{4}R}\ .
\end{align}
We also add $1$ to the denominater of Eq.(\ref{Cuv}) for the case of 
$\mathcal{N}_u\simeq \mathcal{N}_v\simeq 0$.
We further introduce the one dimensional function $C_\textrm{max}(v)$, 
which masures of the constraint violation on the fixed $v$-slice as
\begin{equation}
 C_\textrm{max}(v)=\max_{\textrm{fixed }v} C(u,v)\ .
\end{equation}
Fig.\ref{const_Ndep} shows $C_\textrm{max}(v)$ for $N=4000,8000,16000$. 
(The numerical integration by $N=4000$ broke down at $v\simeq 30$.)
We considered the same setup as Fig.\ref{fig:rcenter}.  
The cutoff near time-like boundaries is fixed as $\delta=0.01$.
The constraint violation keeps small value ($C\textrm{max}\lesssim 10^{-3}$ for $N\gtrsim 8000$).
We can also see $C_\textrm{max}\propto 1/N^2$. 
This is consistent with the fact that our numerical scheme has second order acuracy.

In Fig.\ref{cutoff_dep}, 
we show the time dependence of the tip of the hanging string $r_\textrm{center}(t)$ for several values of the cutoff: $\delta=0.005,0.01,0.02$. The number of grid points are fixed as $N=8000$. 
We again considered the same setup as Fig.\ref{fig:rcenter}. 
The dependence on $\delta$ is small 
and typical chaotic bahaviour of the string does not depend on the value of $\delta$.
Based on the error analysis here, 
we show results in the main text of this paper for $N=8000$ and $\delta=0.01$.

%We now check the dependence of the 
%We introduced the small cutoff $\delta$ near time-like boundories. (See Fig.\ref{fig:grid}.)
%We need to check that our results do not depend on the value of $\delta$.

%\begin{figure}[tb]
%\includegraphics[scale=0.5]{grid_F1_on_D4.eps}
%\caption{\label{const_Ndep} 
%Grid points on the worldsheet for numerical calculations.
%A small cutoff $\delta$ is introduced near time-like boundaries.
%}
%\end{figure}

\section{\label{sec:force}Interquark force from holography}
\subsection{\label{sec:derivation}Derivation}

Here we derive the formula \eqref{eq:force} giving 
a relation between the force acting on quarks in the gauge theory and 
the NG string in the gravity side.
We follow the argument given in App.~D of \cite{Ishii:2015wua}.

We write the on-shell NG action as 
\begin{align}
S[\vec{x}_{q}, \vec{x}_{\bar{q}}] = S_{\text{NG}}[\tilde{X}], 
\end{align}
where $\tilde{X}$ is a solution of the equation of motion with the boundary condition $X(t,r\to \pi/2) = \vec{x}_{q}(t), \vec{x}_{\bar{q}}(t)$. 
%The following argument is totally parallel to that of \cite{Ishii:2015wua}. 
The force acting on quarks in the gauge theory is holographically given by \cite{Ishii:2015wua}
\begin{align}
\langle \vec{F}(t) \rangle = m\partial_{t}(\gamma \vec{v}) + \frac{\delta S[\vec{x}_{q}, \vec{x}_{\bar{q}}]}{\delta\vec{x}_{q}(t)}, 
\end{align}
where $m$ is mass of the quark, $\vec{v}=\dot{\vec{x}}_{q}$ and $\gamma = (1-\vec{v}^{2})^{-1/2}$.

Now, let us evaluate $\delta S/\delta \vec{x}_{q}$ in the gravity side. 
Since the background metric is now given by (\ref{eq:d4metricr}), in the static gauge the NG action becomes
\begin{align}
S_{\text{NG}} &= -\frac{2\lambda M_{\text{KK}}^{2}}{27\pi}\int dtdr \frac{1}{\cos^{3}r} \nonumber \\
& \times \bigg[ \left( 1-\big( \dot{\vec{X}} \big)^{2} \right) \left( \big( \vec{X}' \big)^{2} + \frac{9/M_{\text{KK}}^{2}}{1+\cos^{2}r + \cos^{4}r} \right) \nonumber \\ 
&\qquad+ \big( \dot{\vec{X}}\cdot \vec{X}' \big)^{2}  \bigg]^{-1/2}, \label{eq:ngaction}
\end{align}
where $\dot{} = \partial_{t}$ and $' = \partial_{r}$. 
In what follows we always take our unit $M_{\text{KK}} = 3/2$. 
Solving the equation of motion for $\vec{X}$ near the D4-boundary; $r=\pi/2$, we obtain an asymptotic expansion form of the solution as
\begin{align}
\vec{X}(t,r) = \vec{x}_{q}(t) - \gamma^{2}\vec{a} \epsilon^{2} + \vec{f}_{4}(t) \epsilon^{4} + \mathcal{O}(\epsilon^{5}), \label{eq:assolution}
\end{align}
where $\vec{a}=\ddot{\vec{x}}_{q}$ and we have defined $\epsilon = \pi/2 - r$.

%On one hand, 
To obtain the force, 
let us consider the variation of the action (\ref{eq:ngaction}), 
\begin{align}
\delta S_{\text{NG}} = -\frac{\lambda}{6\pi} \int dt\,  \delta\vec{X}\cdot \left. \frac{\partial \mathcal{L}}{\partial \vec{X}'} \right|_{r=\pi/2-\epsilon}, \label{eq:variation} 
\end{align}
where $\mathcal{L}$ is the integrand of (\ref{eq:ngaction}) and we introduce a cutoff at $r=\pi/2-\epsilon$. 
Substituting the asymptotic solution (\ref{eq:assolution}) into (\ref{eq:variation}), we obtain 
\begin{align}
\delta S[\vec{x}_{q}, \vec{x}_{\bar{q}}] &= \int dt\, \delta \vec{x}_{q}\cdot \bigg[ -\frac{\lambda}{6\pi \epsilon^{2}}\partial_{t}(\gamma \vec{v}) \nonumber \\
&\qquad+\frac{\lambda}{3\pi\gamma}\left( \vec{f}_{4} + \gamma^{2}(\vec{v}\cdot\vec{f}_{4})\vec{v} \right) + A \bigg]. 
\end{align}
The last term $A$ involves complicated terms, but when we consider a probe approximation $\dot{\vec{x}}_{q}\to 0$, $A$ actually vanishes, so we do not care about $A$. 
From this expression we find that the quark mass $m$ corresponds to $\lambda/6\pi\epsilon^{2}$, which is divergent when $\epsilon\to 0$. 
Setting $m=\lambda/6\pi\epsilon^{2}$ and considering probe approximation $\dot{\vec{x}}_{q}\to 0$, we get the force acting on the quark 
\begin{align}
\langle \vec{F}(t) \rangle = \frac{\lambda}{72\pi}\partial_{r}^{4}\vec{X}(t,r)|_{r=\pi/2}, 
\label{eq:F}
\end{align}
where we have replaced $\vec{f}_{4}$ with $\partial_{r}^{4}\vec{X}/4!$.

\subsection{\label{sec:forcesensi}Sensitivity of the interquark force}

To numerically compute the sensitivity of the interquark force to initial perturbations, we can employ two procedures to do it. 
One is a direct calculation: In Eq.~\eqref{eq:F}, change the worldsheet coordinate to double null $u-v$ coordinate and consider linear perturbations $\Phi \to \Phi + \delta \Phi$. 
Evaluating the linearized differential equation for $\delta \Phi=(\delta T, \delta r, \delta X)$ at $r=\pi/2$, we obtain the sensitivity of the interquark force $\langle \delta \vec{F} \rangle$. 
However, this is a little tough since the right hand side of Eq.~\eqref{eq:F} has four derivatives of $r$. 
So, we employ the other one which we explain in the following. 

In Eq.~\eqref{eq:variation}, The integrand is explicitly written as
\begin{align}
\frac{\partial \mathcal{L}}{\partial \vec{X}'} = \frac{1}{\cos^{3}r} \frac{\vec{X}'}{\left[\big(\vec{X}'\big)^2 + 4\big( 1-\big( \dot{\vec{X}} \big)^{2} \big)/\left(1+\cos^{2}r + \cos^{4}r\right)\right]^{\frac{1}{2}}}. 
\label{eq:Fex}
\end{align}
Substituting a solution $\tilde{X}$ to the right hand side and evaluating it at $r=\pi/2$, this quantity is corresponding to infinitely heavy quarkmass and the interquark force. 
Changing to the double null coordinate, 
\begin{align}
\begin{pmatrix}
\vec{X}'  \\
\dot{\vec{X}} \\
\end{pmatrix}
=
\begin{pmatrix}
r_{,u} & T_{,u} \\
r_{,v}  & T_{,v} \\
\end{pmatrix}^{-1}
\begin{pmatrix}
\vec{X}_{,u}  \\
\vec{X}_{,v} \\
\end{pmatrix}
\end{align}
and plugging them into \eqref{eq:Fex}, we find that this can be more easily evaluated numerically since the right hand side includes only single derivatives. 
Considering linear perturbations $\Phi \to \Phi + \delta \Phi$ and evaluating it at $u=0$ or $u=\pi$, we obtain the sensitivity to perturbations of the interquark force: 
\begin{align}
\langle \delta \vec{F}(t) \rangle = \frac{\lambda}{6\pi} \delta \! \left( \frac{\partial \mathcal{L}}{\partial \vec{X}'}  \right) \big|_{u=0,\pi}. 
\end{align}

\nocite{*}

\bibliographystyle{apsrev4-1}
\bibliography{f1chaos}% Produces the bibliography via BibTeX.

\end{document}